\documentclass{aa}
\usepackage{graphicx}
\usepackage{txfonts}

\begin{document}
   \title{Search for exoplanets with the radial-velocity technique: quantitative diagnostics of stellar activity\thanks{Based on observations made with the {\small ESO/HARPS} spectrograph at the 3.6m telescope, La Silla}}

   \author{
     M. Desort \inst{1}
     \and
     A.-M. Lagrange \inst{1}
     \and
     F. Galland \inst{1}
     \and
     S. Udry \inst{2}
     \and
     M. Mayor \inst{2}
   }

   \offprints{
     M. Desort,\\
     \email{Morgan.Desort@obs.ujf-grenoble.fr}
   }

   \institute{
     Laboratoire d'Astrophysique de l'Observatoire de Grenoble, Universit\'e Joseph Fourier, BP 53, 38041 Grenoble, France
     \and
     Observatoire de Gen\`eve, Universit\'e de Gen\`eve, 51 Ch. des Maillettes, 1290 Sauverny, Switzerland
   }

   \date{Received date / Accepted date}

   
   \abstract
   {}
   {It is known that stellar activity may complicate the analysis of high-precision radial-velocity spectroscopic data when looking for exoplanets signatures. We aim at quantifying the impact of stellar spots on stars with various spectral types and rotational velocities and comparing the simulations with data obtained with the {\small HARPS} spectrograph.
   }
   {We have developed detailed simulations of stellar spots and estimated their effects on a number of observables commonly used in the analysis of radial-velocity data when looking for extrasolar planets, such as radial-velocity curves, cross-correlation functions, bisector velocity spans and photometric curves. Stellar and spot properties are taken into account, as well as the characteristics of the spectrograph used (generally {\small HARPS}). The computed stellar spectra are then analyzed in the same way as when searching for exoplanets.
   }
   {1) A first grid of simulation results (radial-velocity amplitudes, bisector velocity-span amplitudes and shapes, and photometry) is built for F-K type stars, with different stellar and spot properties.
   2) It is shown quantitatively that star spots with typical sizes of 1\% can mimic both radial-velocity curves and the bisector behavior of short-period giant planets around G-K type stars with a $v\sin{i}$ lower than the spectrograph resolution. For stars with intermediate $v\sin{i}$, smaller spots may produce similar features. Such spots may complicate the search for low-mass planets on short-period orbits. In these cases, additional observables ({\it e.g.}, photometry, spectroscopic diagnostics) are mandatory to confirm the presence of short-period planets. We discuss these possibilities and show that, in some cases, photometric variations may not be enough to clearly rule out spots as explanations of the observed radial-velocity variations. This is particularly important when searching for super-Earth planets.
3) It is also stressed that quantitative values obtained for radial-velocity and bisector velocity-span amplitudes depend strongly on the detailed star properties, on the spectrograph used, on the line or set of lines used, and on the way they are measured. High-resolution spectrographs will help in distinguishing between spots and planets.}
   {} 
   \keywords{techniques: radial velocities - stars: early-type - stars: planetary systems - stars: starspots - stars: activity}

   \maketitle

\section{Introduction}
Most of the exoplanets detected so far have been identified thanks to the so-called radial velocity (RV) technique, and many more are expected to be found with presently ongoing surveys.
Moreover, thanks to precision improvement, planets with masses as low as a few Earth masses are {\it a priori} detectable, and it is expected that planets with masses similar to the Earth's will be detectable in the future if precision of a few cm\,s$^{\rm -1}$ is reached.
However, it has been known for a long time that stellar activity (spots, pulsations) can also induce RV variations that can be periodic and mimic those induced by planets. Then, there might be risks of misinterpretating RV variations when they have periods shorter than or equal to the star rotational period.

An illustration of potential difficulties encountered with stellar phenomena is \object{51 Peg} itself. \cite{gray97} (1997) argued that the RV variations reported by \cite{mayor95} (1995) and attributed to an orbiting planet with a 4-day period were instead due to some stellar phenomenon. \cite{gray97} (1997) indeed claimed the presence of bisector shapes characteristic of stellar spots or pulsations. However, no quantitative modeling of the amplitude expected for RV or bisector variations were provided. It rapidly turned out that these claims were wrong: in fact \cite{hatzes98} (1998) did not confirm those peculiar bisector variations on the basis of more data at a higher signal-to-noise ratio (SN). Also, very precise photometry did not reveal any variations down to a level of $\Delta(b+y)/2$ = 0.2\,$\pm$ 0.2\,mmag (\cite{henry97} 1997), sufficient enough to reveal the presence of stellar spots. Finally, the planet around \object{51 Peg} was confirmed, even though \cite{brown98} (1998), based on a detailed analysis of the possible effect of pulsations, concluded that they cannot exclude the possibility of pulsations as the source of \object{51 Peg} RV variations. This example shows that stellar activity (spots, inhomogeneities, pulsations) has to be taken into account when analyzing RV data of potentially short-period planets. We focus in the following on star spots.

When modeling stellar activity, \cite{saar97} (1997) made first quantitative estimations of the impact of stellar spots on the RV curve of Fe {\small I} lines at about 6000\,\AA, in the simple case of an equatorial spot with $T_{\rm spot}$ = 0\,K on an edge-on solar-type star.
They showed that peak-to-peak RV amplitudes up to a few hundred m\,s$^{\rm -1}$ can be produced by spots or convective inhomogeneities, depending on the spot size and the projected rotational velocity of the star. They derived quantitative laws for the RV amplitude, as well as for the bisector velocity-span\footnote{The bisector velocity-span measures the global slope of the bisector (\cite{hatzes96} 1996) and allows shape variations of the lines to be detected.} amplitude. They showed that the effect of the star's projected rotational velocity is important for detecting RV variations, but even more for detecting bisector variations.
\cite{saar98} (1998) also estimated the impact of inhomogeneous convection from the study of the bisector velocity-span itself. They conclude that inhomogeneous convection also leads to RV and bisector velocity-span variations up to a few tens m\,s$^{\rm -1}$. They measured the weighted dispersions on RV measurements on a sample of G and F-type stars and found that, for G-type stars with ages about 0.3\,Gyr (with rotational velocities of about 8-10\,km\,s$^{\rm -1}$), $\sigma_v'$ ranges between 20 and 45\,m\,s$^{\rm -1}$. Later on, \cite{paulson04} (2004) measured a jitter up to 50\,m\,s$^{\rm -1}$ due to stellar activity in a sample of Hyades dwarfs. Note that \cite{saar97} (1997) had also pointed out that convective inhomogeneities can lead to even larger RV variations, especially for G2V-type stars.

Generally, observers have tried to examine either the photometric curves or the bisector variations, in addition to RV data, whenever RV variations with short periods were observed. As far as photometry is concerned, the maximum amplitude of variations in milli-magnitude is $\simeq$ 2.5$f$ (\cite{saar97} 1997), where $f$ (in \%) is the fraction of the stellar disk covered with the spot. Depending on spot size and photometric precision, such variations may or may not be detectable. An example of the impact of these complete studies is \object{HD 166\,435} ($\simeq$ 200 Myrs, G0V), for which \cite{queloz01} (2001) rejected one short-period (3.8 days) planet candidate, based on bisector measurements, Ca {\small II} lines, and photometric observations.

In this paper, we investigate in more detail the impact of stellar spots on RV curves and on other diagnostics that are commonly used to disentangle cases of stellar activity from those of planets. We compute the visible spectra of stars with various spectral types (from F to K), projected rotational velocities and orientations, covered with one spot of different sizes and at different positions.
Then, we quantify the resulting RV, bisector velocity-span, and photometric variations. We finally present the results and discuss the impact on RV studies.

  \section{Description of the simulations}

We define the following parameters for the star: temperature $T_{\rm eff}$, gravity $\log g$, rotational velocity $v_{\rm rot}$, inclination {\it i}, metallicity [Fe/H], microturbulent velocity $v_{\rm micro}$, macroturbulent velocity $v_{\rm macro}$. We assume $v_{\rm micro}$ = 1.5\,km\,s$^{\rm -1}$ and $v_{\rm macro}$ = 0.9\,km\,s$^{\rm -1}$ for a G2V star, and a limb-darkening coefficient $\epsilon = 0.6$.
For the spot, the free parameters are the spot colatitude $\theta$ on the star surface, the temperature $T_{\rm spot}$, and the spot size described by the parameter $f_r$, which is the fraction of the visible hemisphere covered by the spot\footnote{For a small spot ($\alpha \ll 1$) defined by its semi-angle $\alpha$ ($2\alpha$ is the angle under which the spot is seen from the star center), $f_r = 1-\cos\alpha$. Note that this fraction is not identical to the fraction of the projected area covered by the spot $f_p$ on the 2D stellar disk, which is equal to $\sin^2\alpha$ for a low spot. For small values of $\alpha$, we have $f_p = 2 f_r$. Hereafter the spot size used is assumed to be $f_r$ unless specified.}.

\subsection{Spectrum computation}

Our simulations use Kurucz models (\cite{kurucz93} 1993). The 3D stellar surface is divided into longitudinal and latitudinal sections. The number of cells is tuned so that the velocity sampling is better than the resolution of the generated Kurucz spectrum. The resolution of Kurucz spectra is chosen to be twice the intrinsic resolution of the instrument we wish to simulate.
 
For a given set of stellar and spot properties, we first compute a synthetic spectrum using Kurucz models without rotational broadening. Then, we apply this spectrum to each cell, shifting the spectrum according to the Doppler law, taking its radial velocity into account. We then sum up the contributions of each cell, weighted by the cell's projected surface and limb-darkening, and taking the presence of a spot into account, when relevant. If a cell is covered by a spot, we use the black-body law for each wavelength of the spectrum to evaluate its weight compared to the same cell without spot ({\it {\it i.e.}}, at $T_{\rm spot} = T_{\rm eff}$). We generate the resulting stellar spectra at different epochs of the star's rotational phase. Typically, we take 20 epochs to cover the phase. Each spectrum is convolved with the instrumental point spread function (PSF). The spectra are computed in the range 377-691\,nm, corresponding to the High-Accuracy Radial-velocity Planet Searcher ({\small HARPS}, \cite{mayor03} 2003) wavelength range or, in some cases, in a range corresponding only  to one order of the spectrograph (see below).

\subsection{RV, bisector, and photometric variations}
We select, for each order, ranges for the spectrum that will be considered to compute the RVs and cross-correlation functions (CCFs). This selection is made so that we include neither broad lines, such as those of H or Ca, nor those from telluric origin.
From these synthetic spectra, we then used the method described in \cite{chelli00} (2000) and \cite{galland05a} (2005a) to derive the RVs, the CCFs, bisectors, as well as bisector velocity-spans for each order of the spectrum or for the global spectrum.
The method for computing the RVs consists in a kind of correlation in Fourier space of each spectrum and a reference spectrum built by summing up all the spectra, which are specific to this star. This method is fairly equivalent to the CCF one for G-K type stars, but is more suited to the case of A and F-type stars with $v\sin{i}$ typically greater than 10-15\,km\,s$^{\rm -1}$. The bisector shape of the lines and the resulting bisector velocity-span are estimated on the CCF in the same way as in \cite{queloz01} (2001). Note that, whenever we wish to compare with real data, it is possible to add noise to the spectra so as to mimic real data as closely as possible. Finally, we also derive the photometric curve over the rotational period at 550\,nm.

In the following we compute the RV, bisector, and photometric variations over a wavelength range narrower than the full {\small HARPS} wavelength range (some orders were not even considered in some cases), and it corresponds to the part of the spectrum that is not affected by atmospheric lines. This choice was made in order to stick as closely as possible to the reduction of actual A-F data, presented for instance in \cite{galland05a} (2005a,b and 2006a,b).

  \subsection{Example}
\begin{figure}[t!]
    \centering
    \includegraphics[width=1\hsize]{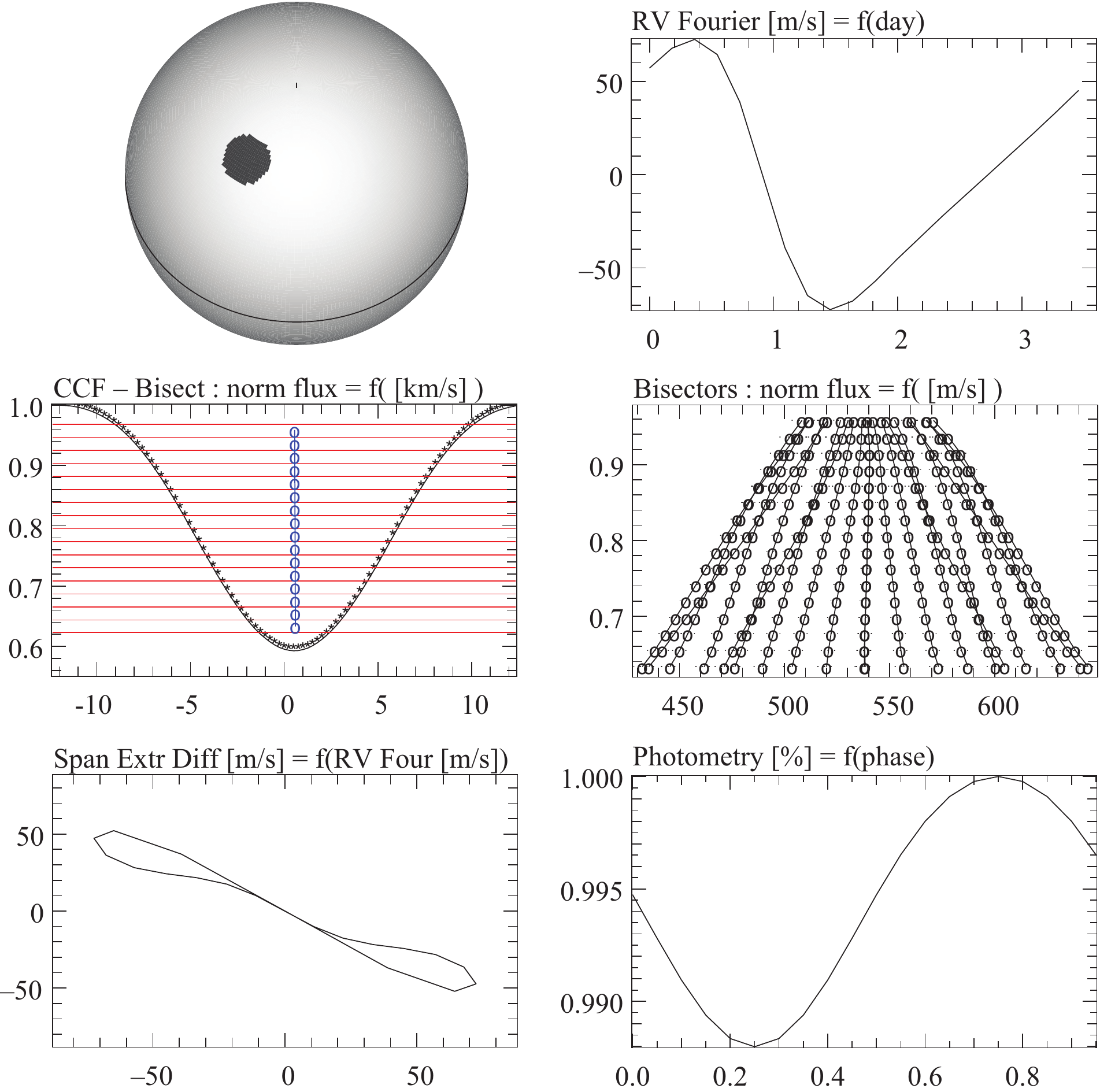}
    \includegraphics[width=0.6\hsize]{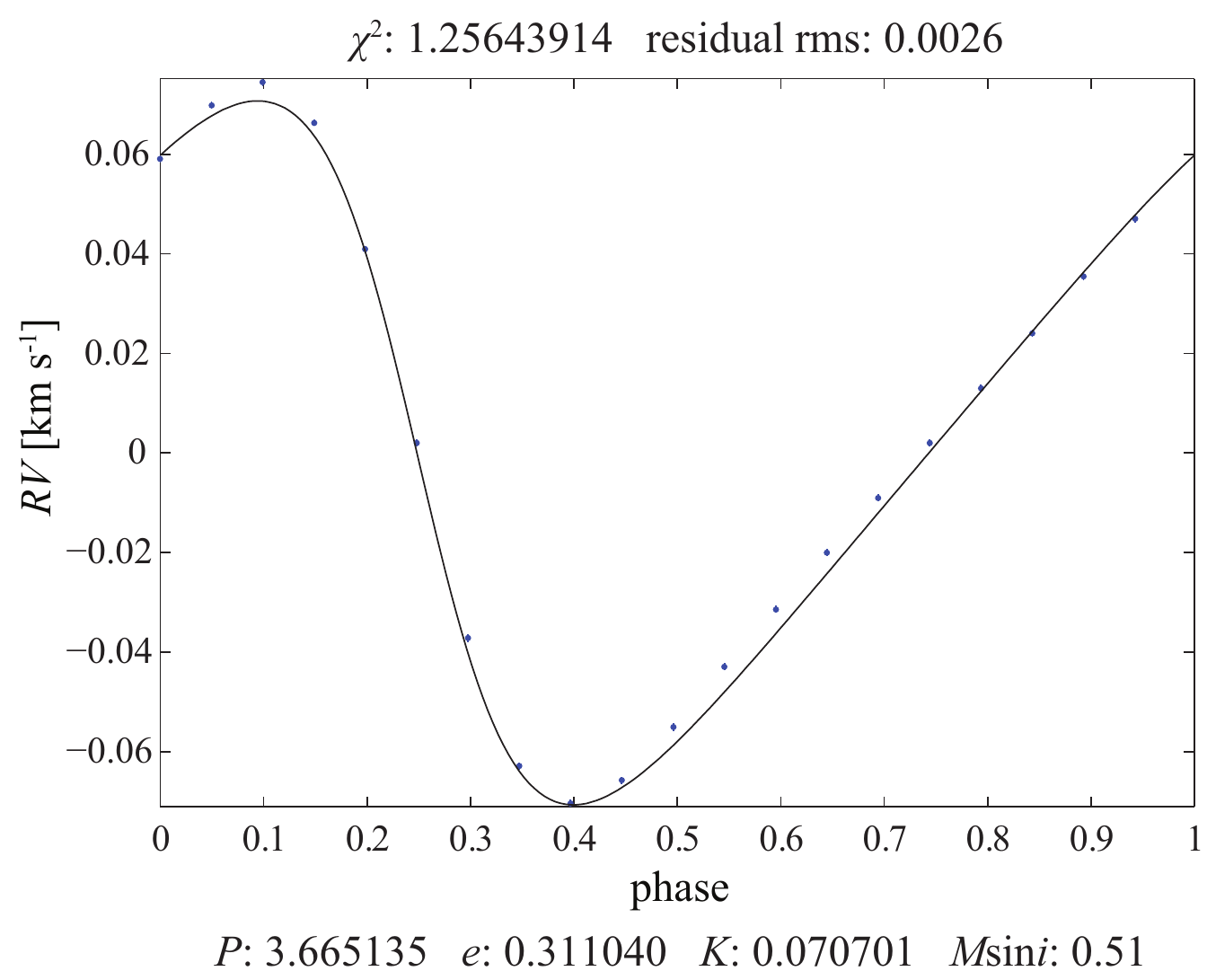}
    \caption{Example of a spot at $\theta$ = 30\degr~with a size of $f_r = 1.02\%$ on a G2V-type star rotating with $v\sin{i}$ = 7\,km\,s$^{\rm -1}$ and seen with an inclination $i$ of 30\degr~with respect to the line of sight: star spot, RV curve, CCFs, bisectors, bisector velocity-span curve, photometric curve. Values measured on the whole spectrum. Part of the spectrum used: orders $\#10$ to $\#58$. A Keplerian model with 0.51\,M$_{\rm Jup}$ and a 3.7-day period would fit the RV data well (bottom).}
    \label{G2_i30_t30_f1p_vsini7km}
  \end{figure}

Figure~\ref{G2_i30_t30_f1p_vsini7km} shows an example of a spot located at a colatitude $\theta$ = 30\degr, covering 1.02\% ($f_r$) of the visible surface of a G2V-type star ($T_{\rm eff}$ = 5800\,K), with $v\sin{i}$ = 7\,km\,s$^{\rm -1}$ and $i$ = 30\degr. The RV curve, CCFs, and bisectors are shown, as well as the bisector velocity-span as a function of RV. Finally, we also show the photometric curve. In this example, we see that the spot produces an RV curve similar to what can be observed when a planet (0.51\,M$_{\rm Jup}$ with a 3.7-day period) orbits a star with a peak-to-peak amplitude of 145\,m\,s$^{\rm -1}$. The bisectors clearly produce an umbrella-like shape, and the bisector velocity-span is clearly correlated to RV variations. Its peak-to-peak amplitude is $\simeq$~120\,m\,s$^{\rm -1}$. The photometric amplitude is 1\% peak-to-peak. This case is similar to the one of \object{HD 166\,435} (\cite{queloz01} 2001).

\section{Order-to-order variations}
\begin{figure}[t!]
    \centering
    \includegraphics[width=0.9\hsize]{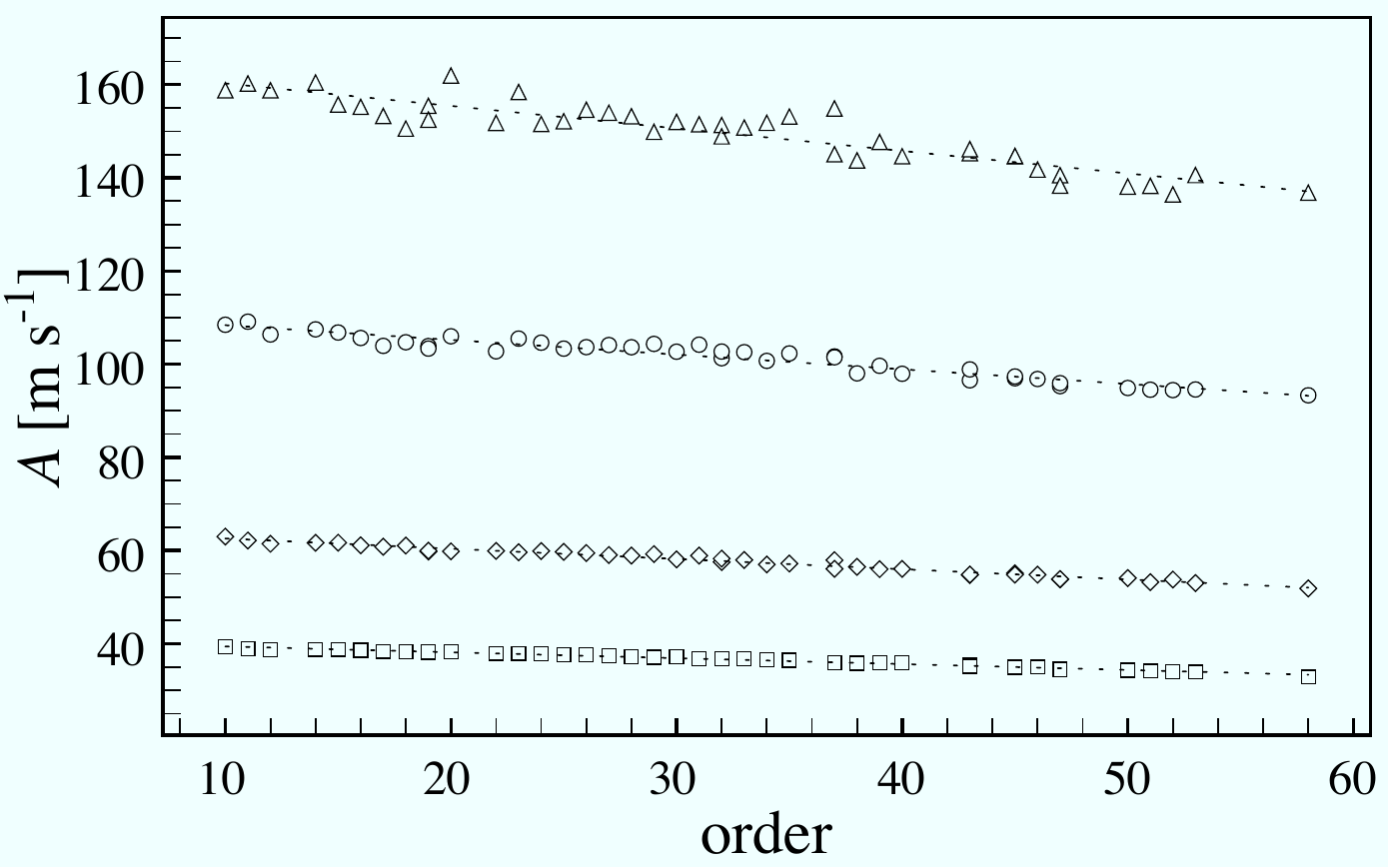}
    \includegraphics[width=0.9\hsize]{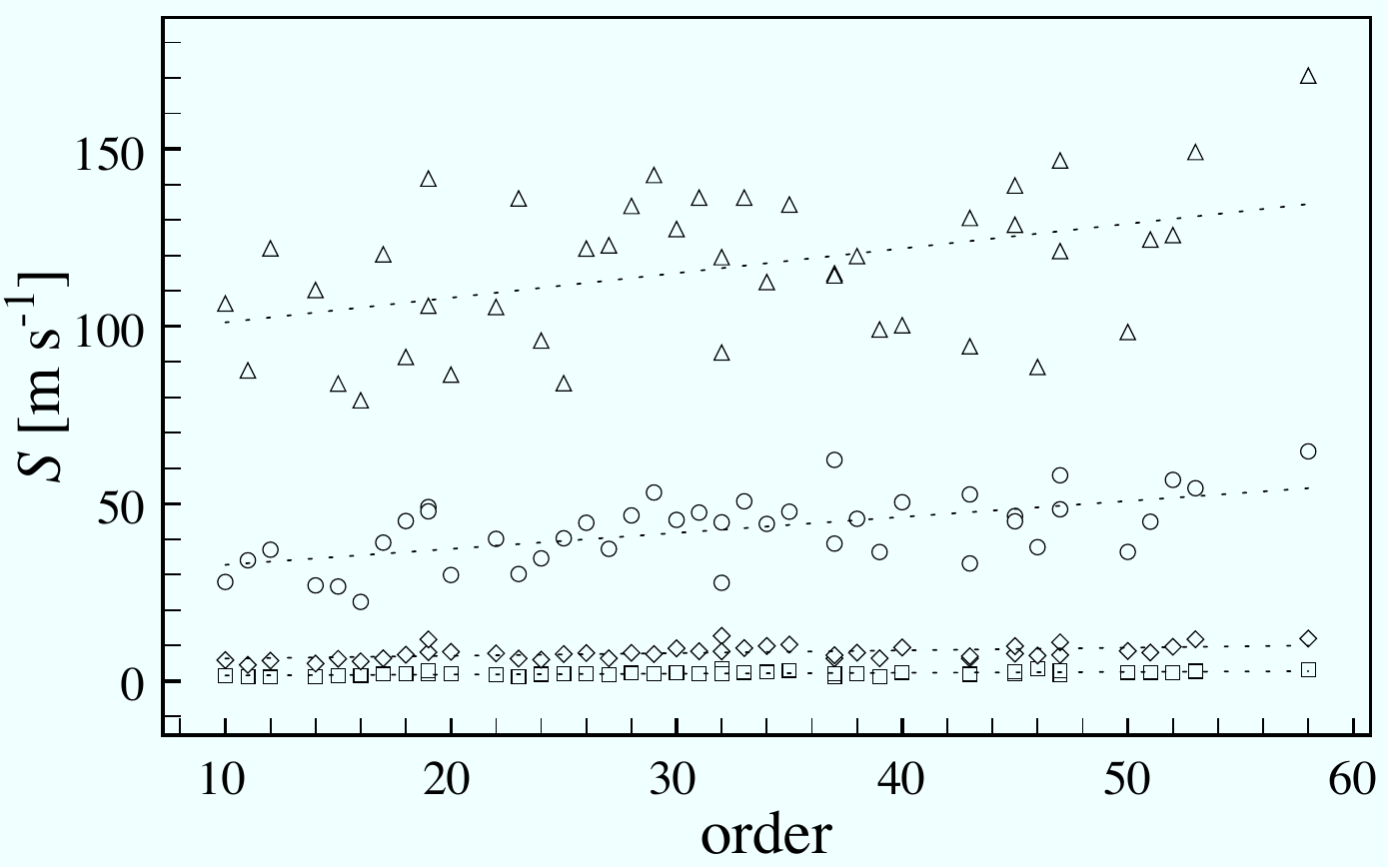}
    \caption{Measured RV amplitude (top) and bisector velocity-span amplitude (bottom) as a function of order number; assuming an equatorial spot with $f_r$ = 1.07\% size on a G2V-type star seen edge-on and with $v\sin{i}$ = 2, 3, 5, and 7\,km\,s$^{\rm -1}$ (from bottom to top curves). Part of the spectrum used: orders $\#10$ to $\#58$.}
    \label{AS_G2_90}
  \end{figure}

We first tried to check the dependence of the obtained values of $A$ (peak-to-peak amplitude of RV variations) and $S$ (peak-to-peak amplitude of the bisector velocity-span variations) on the spectral orders. Figure~\ref{AS_G2_90} shows the values obtained for $A$ and $S$ in the case of $v\sin{i}$ = 2, 3, 5, and 7\,km\,s$^{\rm -1}$ as a function of the order ({\it {\it i.e.}}, of the wavelength). We see that the estimations of $A$ vary within about 10\%, with a smooth trend, decreasing with increasing order. The slopes of the trends for the different $v\sin{i}$ are given in the Table~\ref{linear_fits}. We see that the absolute value of the slope increases with $v\sin{i}$.

The estimations of $S$ show a higher dispersion; nevertheless, a general trend is also observed (see Table~\ref{linear_fits}). This higher dispersion observed for the bisector velocity-span comes from the fact that CCFs and resulting bisector velocity-spans are more sensitive than the RVs to line shapes and blends, given our RV measurement method. As a matter of fact, the CCF is the average of several lines with shapes that differ from one to another; averaging over a single order with a small number of lines (a few tens of lines per order presently) does not completely average out the effects of individual line shapes. This effect is even stronger when considering stars with larger $v\sin{i}$, as more blends, even with faint lines, affect the line shapes. 

There are several consequences:
\begin{itemize}
\item If we are satisfied with results on $A$ with a 10\% range of validity, then we can use only one single order. If we need results with better accuracy, we have to compute the whole stellar spectrum.
\item Given the observed dispersion on the bisector velocity-span, quantitative results or laws derived from a single line or set of lines ({\it {\it e.g.}}, \cite{saar97} 1997 or \cite{hatzes02} 2002) should not be used for quantitative comparison with ``global" CCFs ({\it {\it i.e.}}, obtained using the whole spectrum).
\item Inter-comparison of $S$ values based on averaged CCFs over the whole spectral range should be restricted to stars with similar spectral types and projected rotational velocities because of their strong effect on the broadening of lines.
\item The chromatic dependence of $A$, even though limited to a level of about 10\%, may be used as a very precious additional criterion for spot\,/\,planet diagnostics. This will be developed more later. Also, observing in near-infrared will allow reduction of the effect of spots and thus allow detection of less massive companions of active stars.
\end{itemize}
Note that, even though these trends have been shown in the case of an edge-on solar-type star, they would qualitatively remain for other geometries.

\section{Edge-on G2V-type star -- equatorial spot}

We have so far computed a limited series of simulations of complete spectra for different values of star's projected rotational velocities, inclinations, and different spot characteristics (see Table~\ref{full_cases}). In this section we present the results for an edge-on star ($i$ = 90\degr) with an equatorial spot ($\theta$ = 90\degr).

\begin{table}[ht!]
    \caption{Cases for which full spectra were used (G2V-type stars).}
    \label{full_cases}
    \begin{center}
      \begin{tabular}{l l l l}
        \hline
	\hline
        $i$			& $\theta$			& $f_r$				& $v\sin{i}$\\
        (\degr)		& (\degr)		& (\%)				& (km\,s$^{\rm -1}$)\\
	\hline
    	90           		& 90            		& 1.07				& 2, 3, 5, 7, 10   \\
      	90           		& 90            		& 0.5, 1.07, 1.22, 1.98	& 7   \\
      	60           		& 60            		& 0.99			   	& 2, 5, 7   \\
	30          		& 40            		& 0.99				& 2, 7   \\
	10           		& 60            		& 1.03				& 2, 7   \\
	30          		& 30            		& 0.99				& 2, 5, 7   \\
	10           		& 10            		& 0.95				& 2, 5, 7   \\
   \hline
      \end{tabular}
    \end{center}
\end{table}

\subsection{General results}

\begin{figure}[t!]
    \centering
    \includegraphics[width=1\hsize]{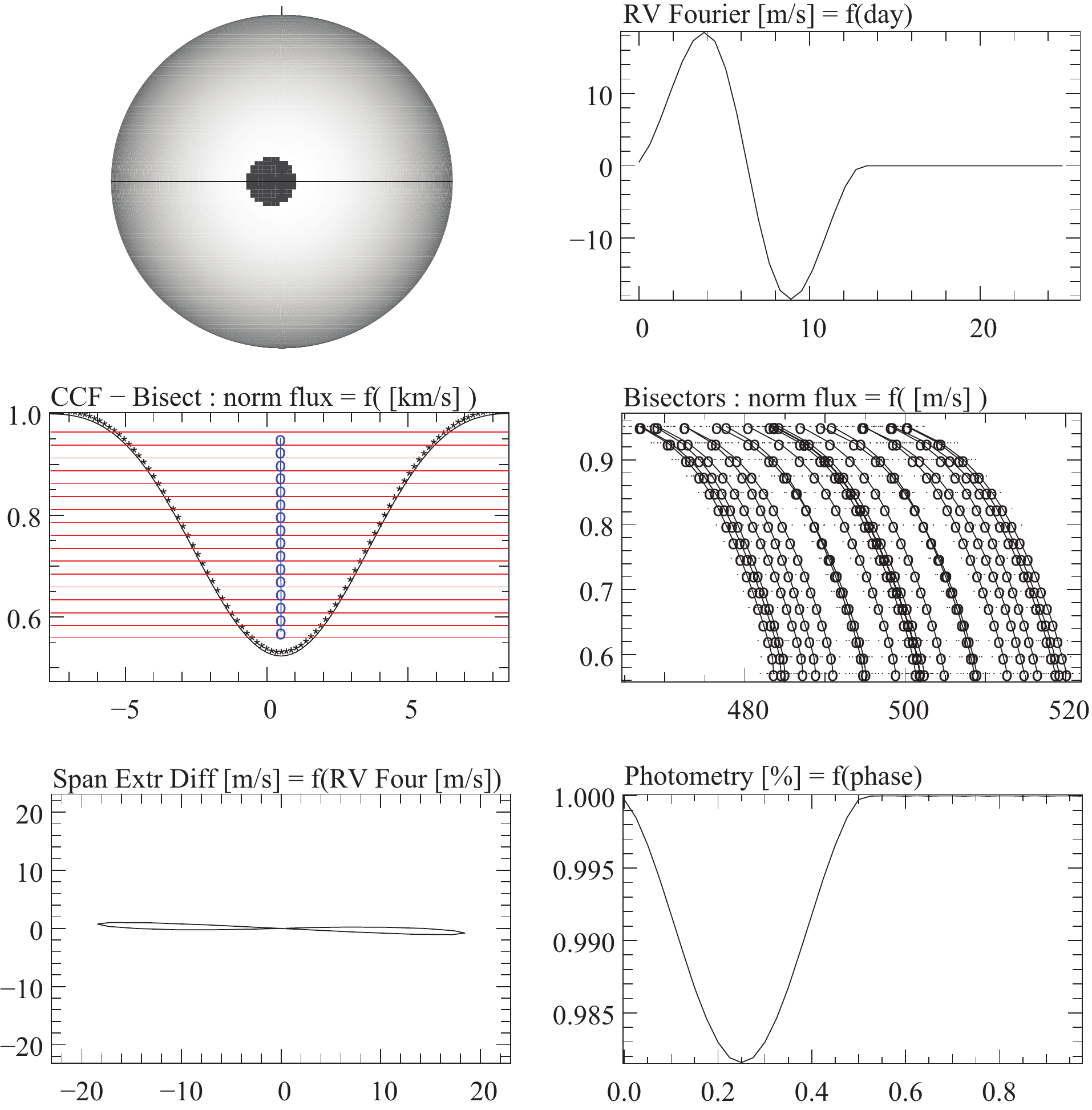}
    \caption{A 1.07\% equatorial spot on an edge-on G2V-type star rotating with $v\sin{i}$ = 2\,km\,s$^{\rm -1}$: star spot, RV curve, CCFs, bisectors, bisector velocity-span curve, photometric curve. Values measured on the whole spectrum. Part of the spectrum used: orders $\#10$ to $\#58$. In this case with a $v\sin{i}$ value lower than the spectrograph resolution, the bisectors are only shifted (without change of shape) even if the origin of RV variations is the presence of a spot.}
    \label{G2_i90_t90_f1p_vsini2km}
  \end{figure}

\begin{figure}[t!]
    \centering
    \includegraphics[width=1\hsize]{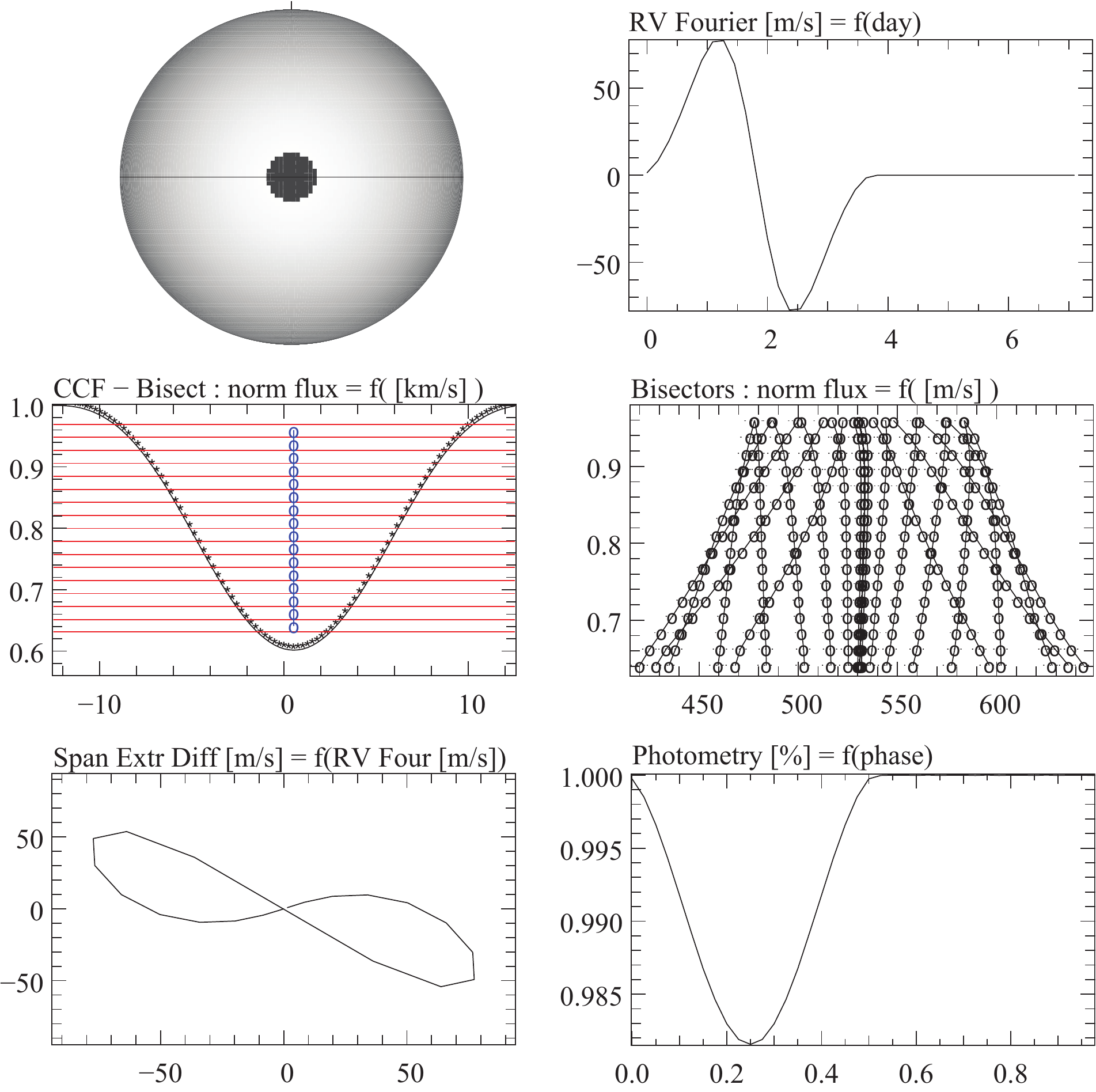}
    \caption{Same as Fig.~\ref{G2_i90_t90_f1p_vsini2km} but with $v\sin{i}$ = 7\,km\,s$^{\rm -1}$. In this case with a $v\sin{i}$ value greater than the spectrograph resolution, the bisectors undergo changes of shape from the presence of a spot that also leads to RV variations.}
    \label{G2_i90_t90_f1p_vsini7km}
  \end{figure}

For a given spot location, the values of $A$, $S$, and the bisector shapes strongly depend on the star's projected rotational velocity. Figures~\ref{G2_i90_t90_f1p_vsini2km} and \ref{G2_i90_t90_f1p_vsini7km} show as an illustration the RV, bisectors, bisector velocity-span, and photometric curves obtained respectively for $v\sin{i}$ = 2 and 7\,km\,s$^{\rm -1}$. The peak-to-peak amplitude of the RV increases from 37 to 155\,m\,s$^{\rm -1}$, {\it {\it i.e.}}, roughly proportional to $v\sin{i}$. This was expected from the first-order simulations quoted above. More striking and important are the differences observed in the bisector variations. Even though we observe an umbrella-like shape for the bisectors in the case of high-enough $v\sin{i}$ (note that this umbrella-like shape is also observed in Fig.~\ref{G2_i30_t30_f1p_vsini7km}), as well as a clear correlation between $A$ and $S$ (inclined ``8'' shape), we only observe a shift in the case of low $v\sin{i}$.
In this case, when the $v\sin{i}$ value becomes lower than the spectrograph resolution (non-resolved lines), the effect of a change of shape in the stellar lines only results in a shift (at first order) after convolution with the instrumental PSF, which one can understand easily: the shape then results essentially from the shape of the PSF. 
In such a case, it will not be possible to find correlations between the bisector velocity spans and RVs.
We estimate that the astrometric effect of this 1\% equatorial spot on an edge-on star at 550\,nm is a photocenter shift of 73\,$\mu$AU (thus, 7.3\,$\mu$as at 10\,pc), assuming a stellar radius equals 1\,R$_{\sun}$.

\subsection{Impact of spot size and $v\sin{i}$ on the RV amplitude}

\begin{figure}[t!]
   \centering
   \includegraphics[width=0.9\hsize]{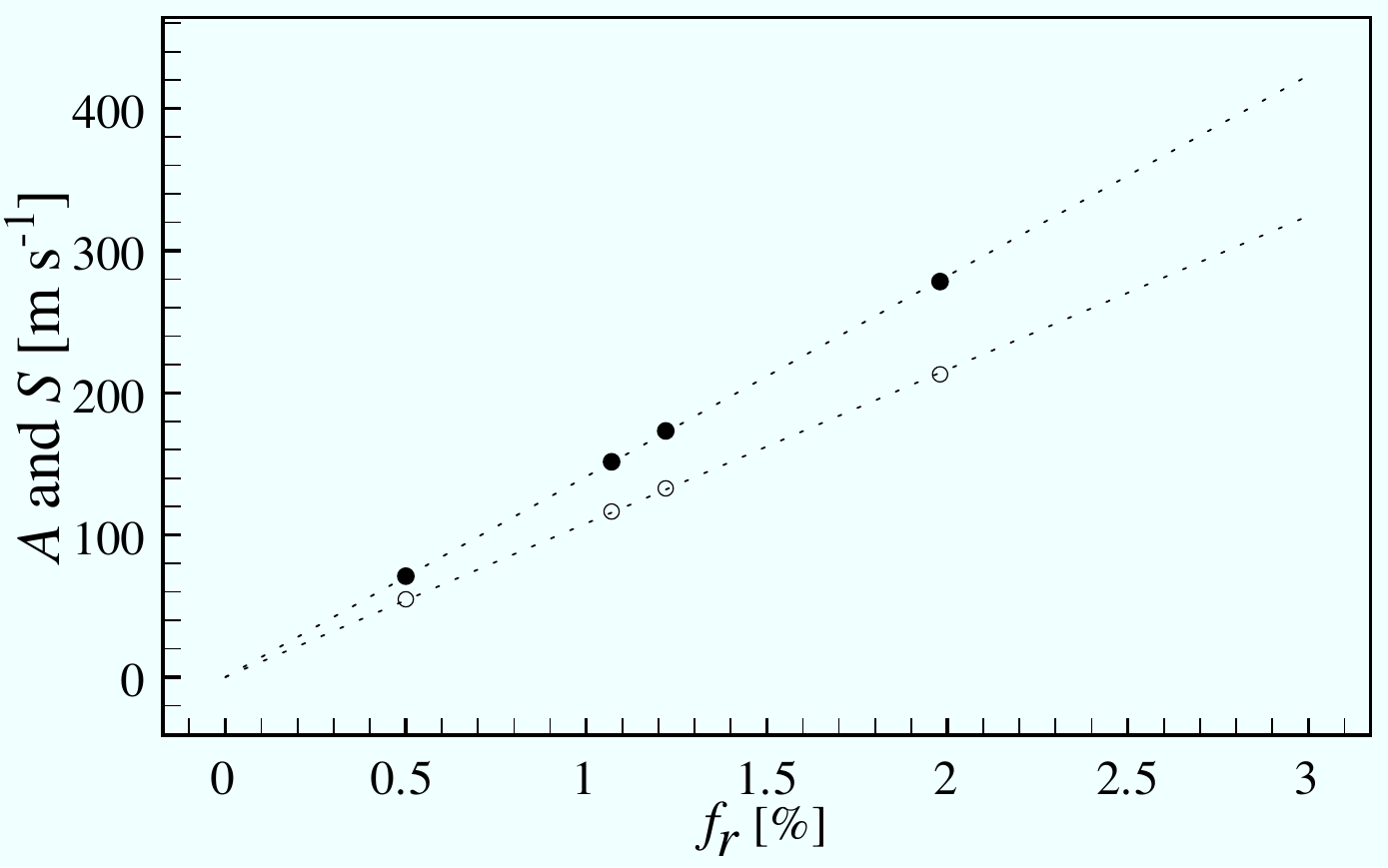}
\caption{$A$ (filled circles) and $S$ (open circles) as a function of the spot size $f_r$, for an equatorial spot on an edge-on G2V-type star with $v\sin{i}$ = 7\,km\,s$^{\rm -1}$. Values measured on the whole spectrum. Part of the spectrum used: orders $\#10$ to $\#58$.}
   \label{G2_i90_t90_fr1_AandS_fr}
  \end{figure}

We first checked with a full spectrum that the effect of the spot size is the same for every order. To do so, we compared the $A$ values obtained for each order, assuming a star with edge-on spots with sizes of 0.5 and 1.07\%. The values are found to be identical from one order to the next one within less than 1\%. We then performed several simulations on order $\#31$, assuming different values of $f_r$ ranging between 0.5 and 2\%. The amplitudes of RV variations measured on order $\#31$ are found to be proportional to the spot size (Fig.~\ref{G2_i90_t90_fr1_AandS_fr}). Altogether, this shows that the RV amplitude measured on the whole spectrum is proportional to the spot size.

\begin{figure}[t!]
   \centering
   \includegraphics[width=0.7\hsize]{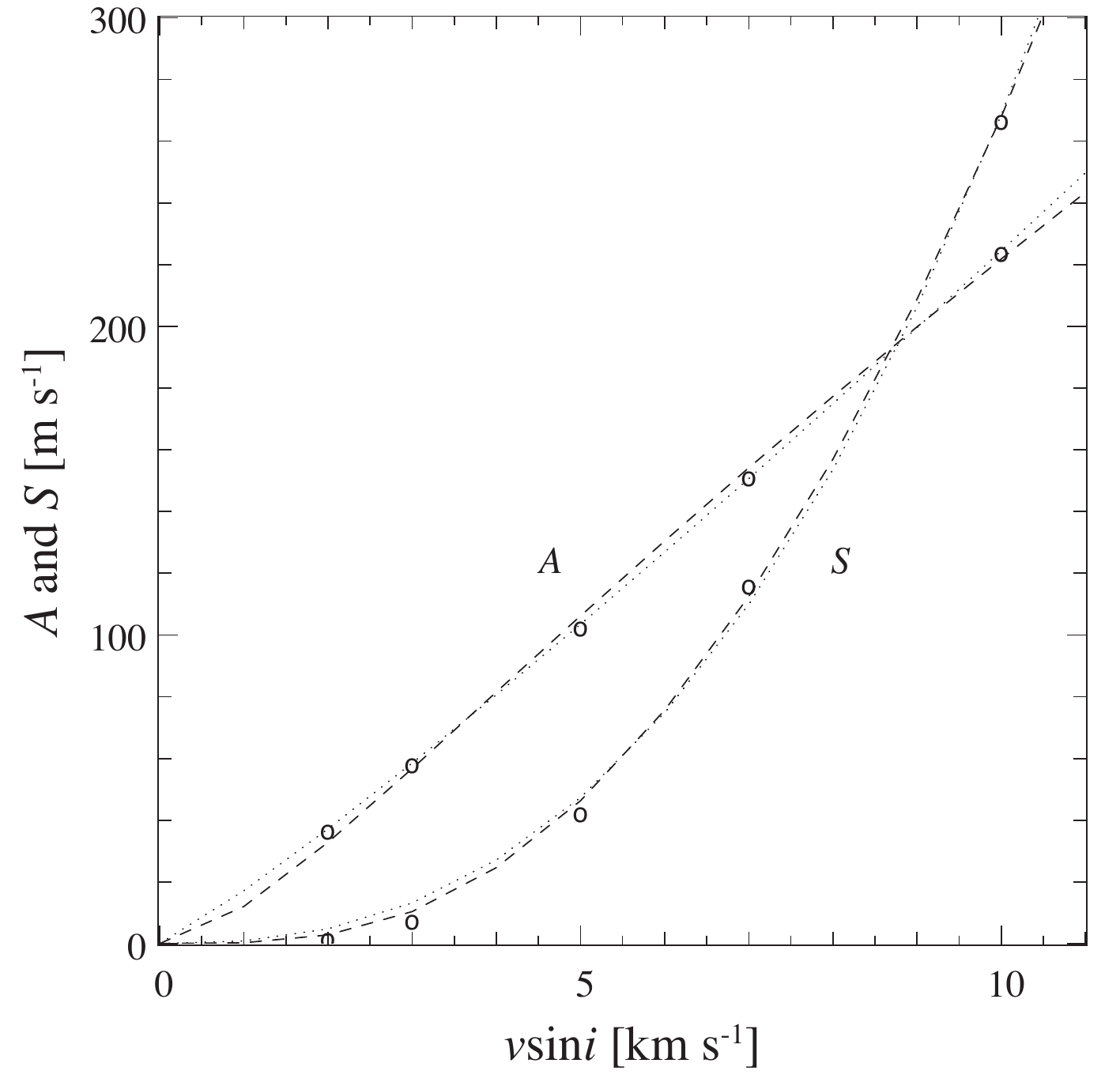}
\caption{$A$ and $S$ as a function of the star's projected rotational velocity, for an equatorial spot on an edge-on G2V-type star. Values measured on the whole spectrum. Part of the spectrum used: orders $\#10$ to $\#58$. $A$ varies approximately linearly with $v\sin{i}$. $S$ varies as $\left(\sqrt{(v\sin{i})^2+v_0^2} - v_0 \right)^{\alpha}$. If we take into account the instrumental resolution (dashed line) ${\alpha} \simeq$ 1.8, otherwise (dotted line) ${\alpha} \simeq$ 2.5 (see text).}
    \label{G2_i90_t90_fp5to2_AandSoverf_vsini}
  \end{figure}

Figure~\ref{G2_i90_t90_fp5to2_AandSoverf_vsini} provides the amplitude of the RV variations, $A$, as a function of $v\sin{i}$, measured on whole spectra, for a spot size of 1\%. We find that in the investigated range of $v\sin{i}$, $A$ depends almost linearly on the star's projected rotational velocity. More precisely, with $f_r$ fixed to 1.07\%, we find $A = 17\,(v\sin{i})^{1.1}$. Then we can conclude that

\begin{equation}
\label{A_G2}
$$A = 16\,f_r\,(v\sin{i})^{1.1}$$,
\end{equation}
where $f_r$ is expressed in percent, and $v\sin{i}$ in km\,s$^{\rm -1}$.

 \subsection{Impact of spot size and $v\sin{i}$ on the bisector velocity-span}
 
Similarly, we checked that the $S/f_r$ values do not depend on the spot size. Then, using spectra over the whole order range, we find that in the investigated range of values of $v\sin{i}$ (2-10\,km\,s$^{\rm -1}$), $S$ strongly depends on the star's projected rotational velocity, as can be seen in Fig.~\ref{G2_i90_t90_fp5to2_AandSoverf_vsini}. More precisely, we find that
\begin{equation}
\label{S_G2}
$$S = 6.5\,f_r\,\left(\sqrt{(v\sin{i})^2+v_0^2} - v_0 \right)^{1.8}$$,
\end{equation}
where $v_0$ is the instrumental width ($v_0$ fixed to 3\,km\,s$^{\rm -1}$ given a resolution of 100\,000 in the case of {\small HARPS}). The resolution plays a crucial role for the bisector determination: if we do not take it into account, we obtain $S = 0.79\,f_r\,(v\sin{i})^{2.5}$, and the $\chi^2$ value for the fit is multiplied by 3. This means that the PSF affects the shape of the star spectrum, and the less the lines are resolved the stronger this effect. If $v\sin{i}$ is negligible in comparison to $v_0$ (lines not resolved), then $S$ is negligible (in comparison to $A$ for example), which is consistent with the fact that the deformation of the stellar spectrum then only results in a small shift of the observed spectrum.

Besides, we note that for a given $v\sin{i}$, $A/S$ does not depend on $f_r$. Hence $A/S$ depends almost only on the star parameters, $T_{\rm spot}$, and spot location.

\subsection{Impact of the instrumental resolution on the bisector velocity-span}

\begin{figure}[t!]
   \centering 
   \includegraphics[width=0.9\hsize]{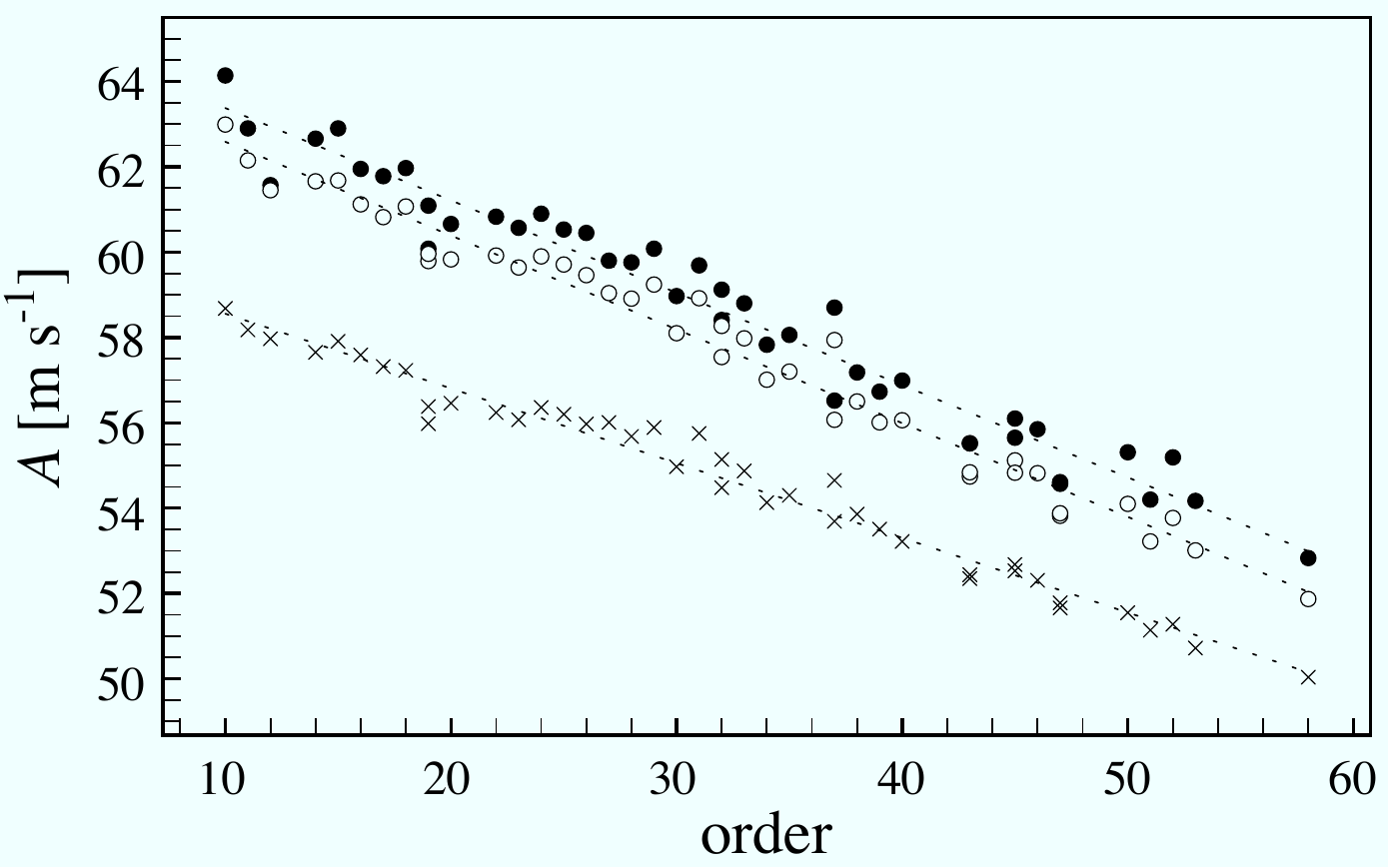}
   \includegraphics[width=0.9\hsize]{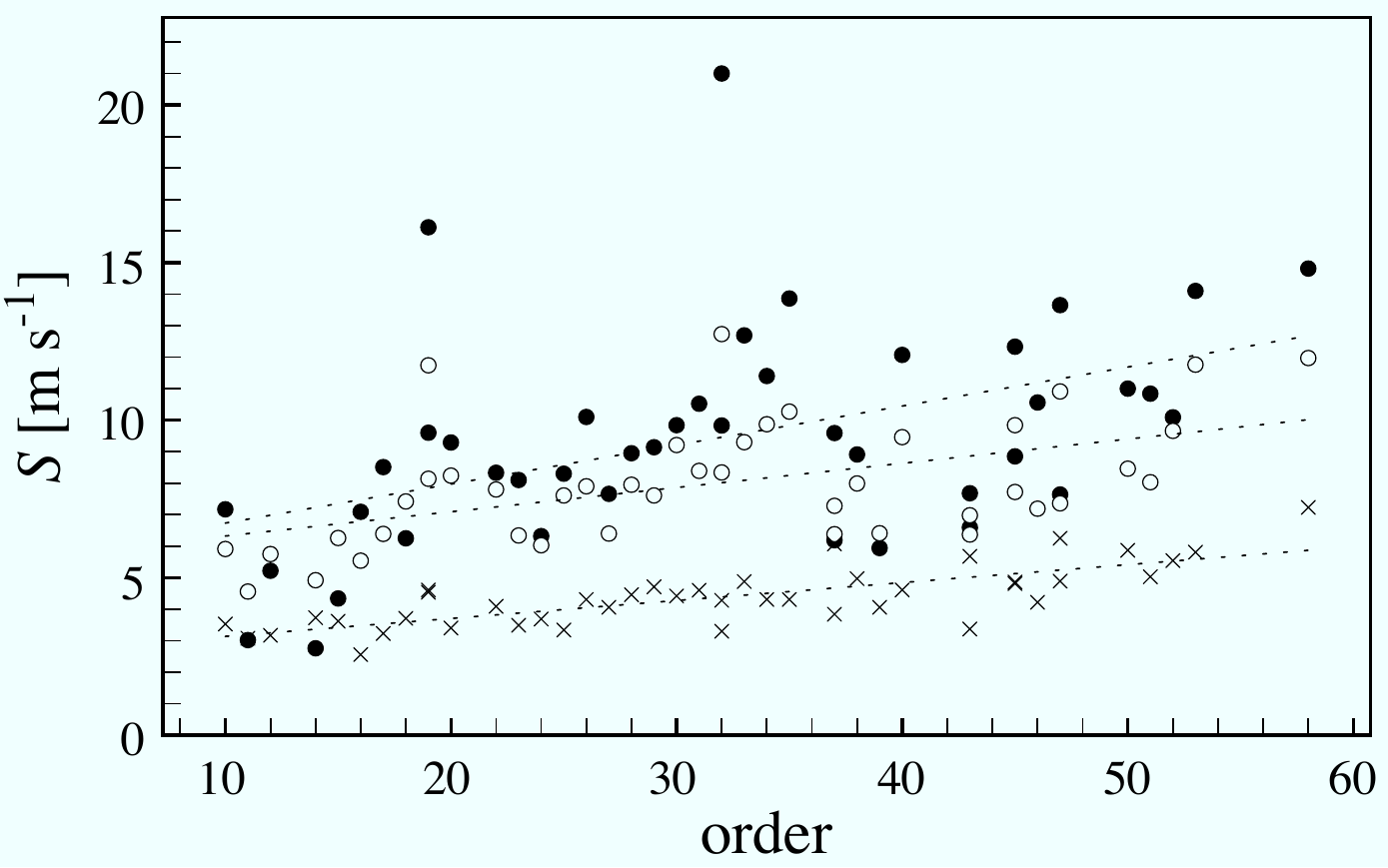}
   \includegraphics[width=0.9\hsize]{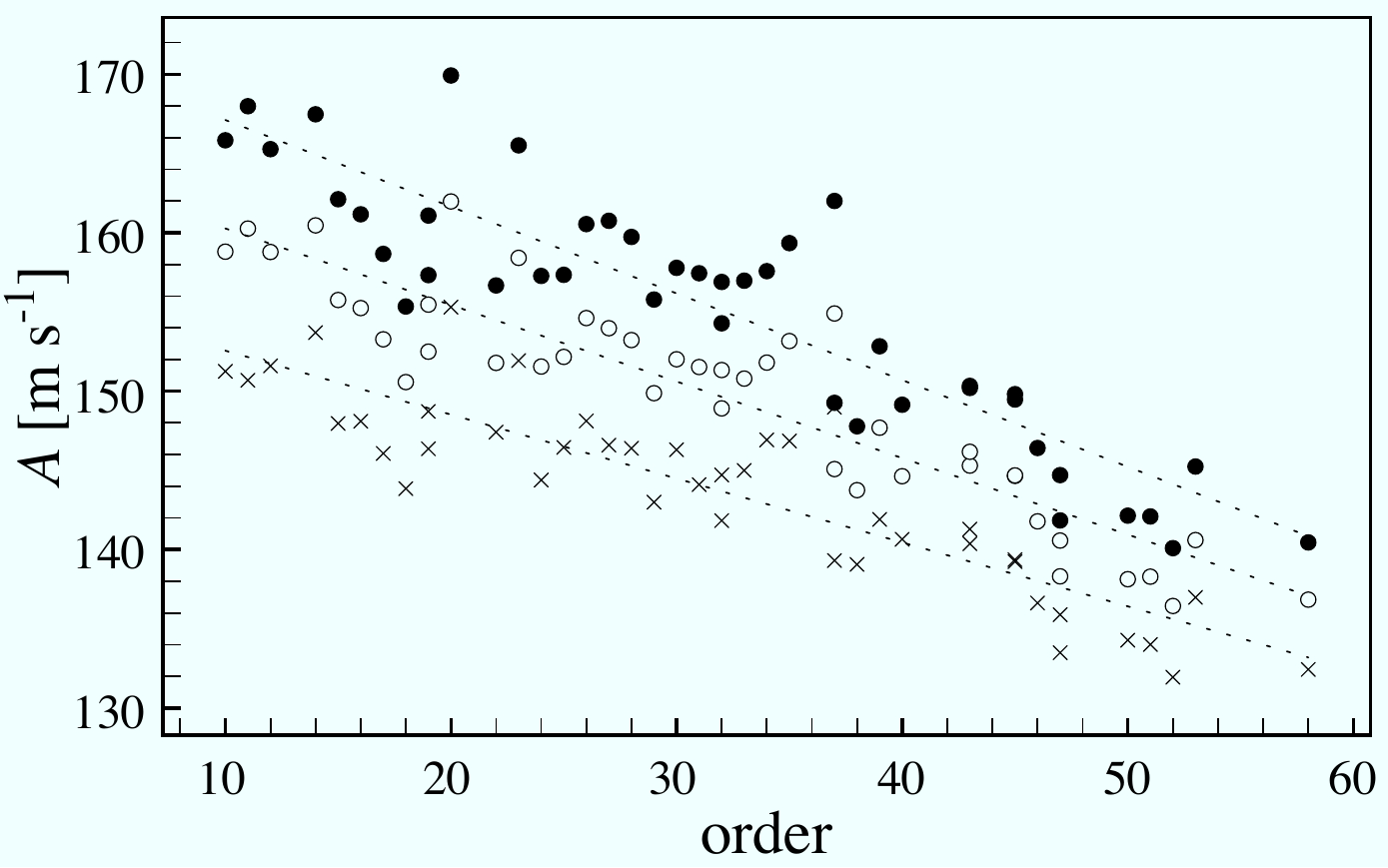}
   \includegraphics[width=0.9\hsize]{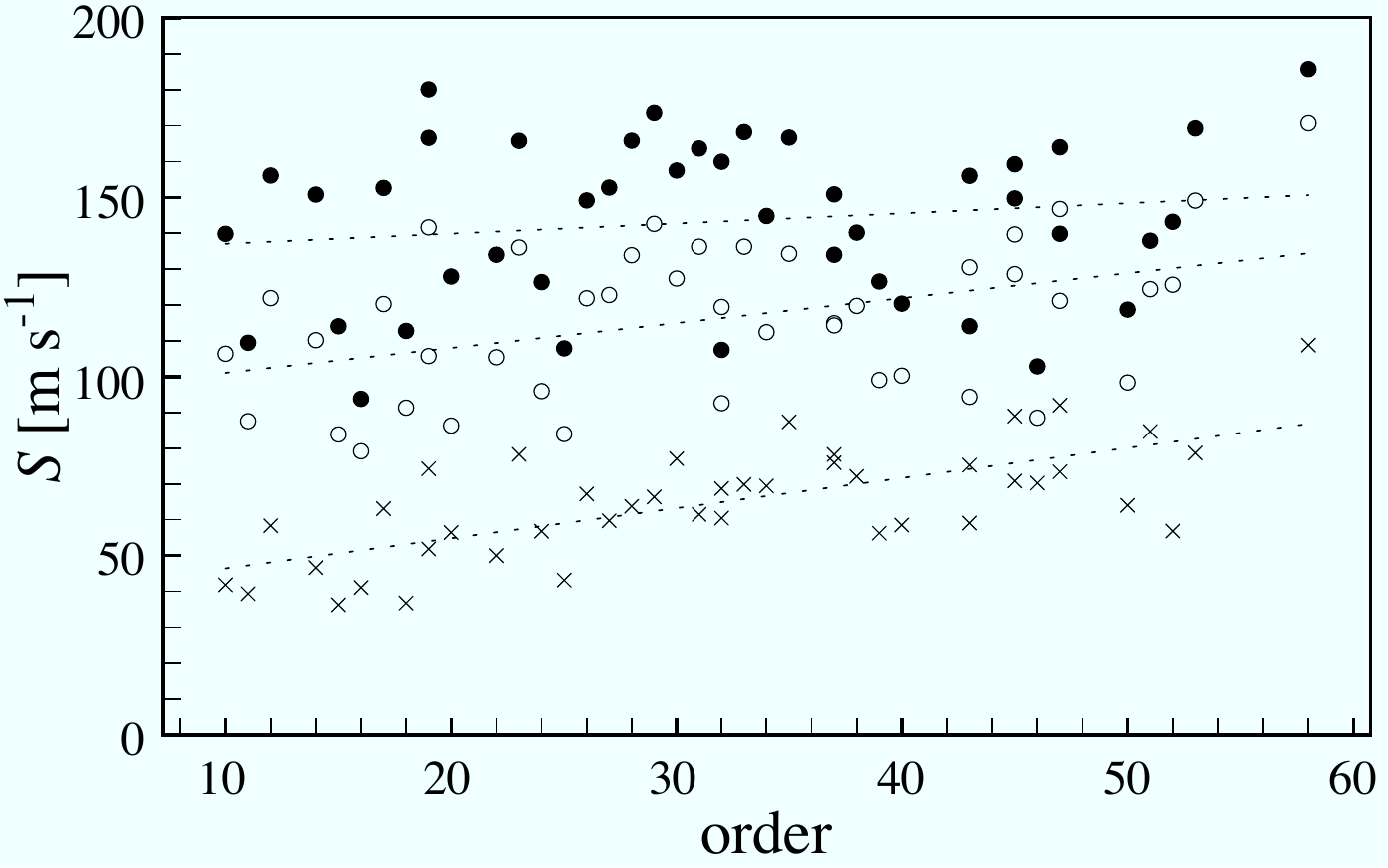}
\caption{$A$ and $S$ as a function of order number when taking the instrumental profile into account (crosses: $R$ = 50\,000; open circles: $R$ = 100\,000) or not (filled circles, corresponding to $R$ = 300\,000). Two cases are shown: G2V-type star with $v\sin{i}$ = 3\,km\,s$^{\rm -1}$ (2 upper panels) and with $v\sin{i}$ = 7\,km\,s$^{\rm -1}$ (2 lower panels). In both cases, an equatorial spot with 1\% size is assumed. Values measured from order $\#10$ to $\#58$.}
    \label{Results_psf_opo_AandS_N_v}
  \end{figure}

To test the impact of instrumental resolution, we computed simulations of an equatorial spot ($f_r$ = 1\%) on a G2V-type star, seen edge-on, taking the instrumental profile into account or not\footnote{Note that even without taking the instrumental profile into account, the spectrum is still sampled over 4096 pixels per order, resulting in an instrumental resolution of 300\,000.}. When taking it into account, we assumed the resolution to be 100\,000 ({\small HARPS}) or 50\,000. Two cases of stellar projected rotational velocities have been considered: 3\,km\,s$^{\rm -1}$ and 7\,km\,s$^{\rm -1}$. Figure~\ref{Results_psf_opo_AandS_N_v} shows the obtained values of $A$ and $S$, order per order. It can be seen that taking the instrumental PSF into account leads to lower values of both $A$ and $S$, and the values decrease when the spectral resolution decreases. We also see that the impact on $S$ is much stronger than the impact on $A$. In the case of the bisector velocity-span, the values obtained are a factor of two smaller than the ones obtained when neglecting the instrumental profile.

\subsection{Comparison with previous results}

Assuming an equatorial spot with a temperature of 0\,K, on an edge-on, solar-type star, \cite{saar97} (1997) find that the amplitude of the RV variations of the 6000\,\AA~Fe {\small I} line, expressed in m\,s$^{\rm -1}$, follows the law $A_S \approx 6.5 f^{\,0.9} v\sin{i}$ where $A_S = A/2$, $f$ is the fractional projected spot size (previously defined $f_p$), and $v\sin{i}$ the projected rotational velocity, expressed in km\,s$^{\rm -1}$. Note that they did not take any instrumental resolution into account. They also derive the following law for the amplitude of bisector velocity-span variations, in m\,s$^{\rm -1}$:
$A_{S,{\rm span}} \sim 0.11 f^{\,0.9} (v\sin{i})^{3.3}$ with $A_{S,{\rm span}} = S/2$.

Using the Ca {\small I} 6439\,\AA~line, and also assuming an equatorial spot with $T_{\rm eff} - T_{\rm spot}$ = 1200\,K on an edge-on solar-type star and a spectral resolution of 0.043\,\AA~($R$ = 150\,000), \cite{hatzes02} (2002) find  $A_S = (8.6V-1.6) f^{\,0.9}$ and $A_{S,{\rm span}} = (22-16V+3.3V^2) f^{\,0.9}$ where $f$ seems to be the projected spot size ($f_p$). He claims that his results are compatible with those of \cite{saar97} (1997), while assuming a very different temperature for the spot. This is in fact surprising as spot temperature certainly impacts the RV and bisector velocity-span amplitudes (see below).
We find the laws (\ref{A_G2}) and (\ref{S_G2}) for an equatorial spot on an edge-on G2V-type star, simulated with {\small HARPS}. They are obtained using the whole spectrum between orders $\#10$ and $\#58$. We recall that $f_r$ = $f_p$/2.

One has to be careful when comparing the results of different simulations. Indeed, as illustrated before, 1) the choice of different lines or wavelength ranges to measure the different parameters (already highlighted by \cite{gray82} 1982), 2) the spot temperatures, and 3) the instrumental resolution, all have a significant impact on the values obtained for the bisector velocity-span.
We find a quasi linear dependence of $A$ as a function of $f$ (Fig.~\ref{G2_i90_t90_fr1_AandS_fr}), as found by the other authors and believe that the differences found between the power dependence on $f$ or $v\sin{i}$ for the amplitude (power 0.9 for \cite{hatzes02} 2002 and power 1 for the present work) is probably not significant.

Considering an edge-on G2V-type star with $v\sin{i}$ = 7\,km\,s$^{\rm -1}$ and $T_{\rm eff}$ = 5800\,K, an equatorial spot with $f_r$ = 1.07\% ($f_p$ = 2.14\%), and $T_{\rm spot}$ = 100\,K, for a single line at 6006\,\AA, we find $A$ = 200\,m\,s$^{\rm -1}$ and $S$ = 238\,m\,s$^{\rm -1}$. \cite{saar97} (1997) find $A$ = 180\,m\,s$^{\rm -1}$ and $S$ = 268\,m\,s$^{\rm -1}$ for $T_{\rm spot}$ = 0\,K, and the results are in reasonable agreement. If we take another line at 6439\,\AA~with $T_{\rm spot}$ = 100\,K, we find $A$ = 192\,m\,s$^{\rm -1}$ and $S$ = 125\,m\,s$^{\rm -1}$. This shows that $S$ strongly depends on the line chosen.
If we now try to compare our results with those of \cite{hatzes02} (2002) with the same parameters, except $T_{\rm spot}$ = 4600\,K and a single line at 6439\,\AA, we find $A$ = 121\,m\,s$^{\rm -1}$ and $S$ = 81\,m\,s$^{\rm -1}$, while he finds $A$ = 232\,m\,s$^{\rm -1}$ and $S$ = 284\,m\,s$^{\rm -1}$. The discrepancy is quite important there. It shows that probably the \cite{saar97} (1997) and \cite{hatzes02} (2002) results were in fact not compatible. We note that using \cite{hatzes02} (2002) laws with $f = f_r$ instead of $f_p$ would lead to more similar results ($A$ = 125\,m\,s$^{\rm -1}$ and $S$ = 152\,m\,s$^{\rm -1}$).

All this shows that one should not make any quantitative comparison of bisector velocity-span values or $A/S$ obtained on real data with formulas such as those proposed by \cite{saar97} (1997), \cite{hatzes02} (2002), or this paper. Dedicated simulations corresponding to the same instrument and star properties should be performed, at least to calibrate the dependence for a given instrument.

\section{Spot at different latitudes on an edge-on G2V-type star}

We ran several single-order ($\#31$) simulations for a spot of various sizes ($f_r$ = 0.5-2\%), located at different colatitudes: $\theta$ = 90, 60, 30, and 10\degr. We find again that, for a given star, the $A/S$ ratio does not depend on the spot size and that $A/f_r$ and $S/f_r$ follow laws as a function of $v\sin{i}$, with powers similar to those obtained in the case of an edge-on star with a similar velocity.

Globally, the amplitudes of $A$ and $S$ decrease with increasing latitudes of the spot. In those circumstances, more cases occur where bisector velocity-span variations may not be detectable. Of course, in the present case (edge-on star), the RV curve and the photometric curve will be constant half of the time, and it will be easier to distinguish RV variations due to a companion from RV variations due to a stellar spot. Nevertheless, cases with several spots might be problematic.


\section{Spot on an inclined G2V-type star}

Obviously this case is more interesting in the framework of RV planets as the spot may be seen over the whole rotational phase of the star. As before, we computed a few simulations over the whole spectrum (see Table~\ref{full_cases}), as well as more numerous single-order ($\#31$) simulations of a spot on an inclined star. For the detailed simulations, we assumed a spot of 1\%.

\subsection{Examples}

\begin{figure}[t!]
    \centering
    \includegraphics[width=1\hsize]{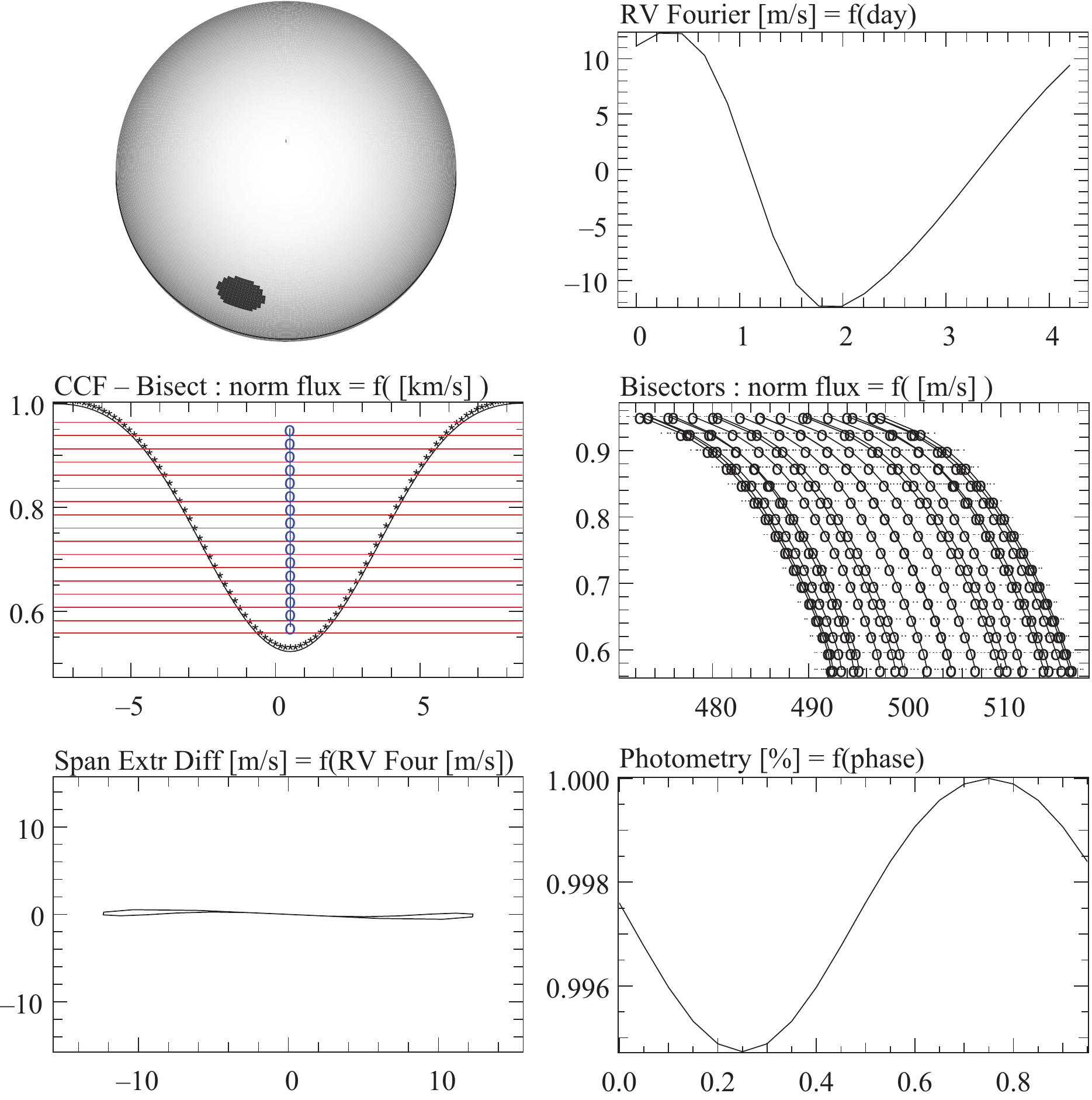}
    \caption{Spot located at $\theta$ = 60\degr~with a size of 1.03\% on a G2V-type star seen almost pole-on ($i$ = 10\degr), rotating with $v\sin{i}$ = 2\,km\,s$^{\rm -1}$: star spot, RV curve, CCFs, bisectors, bisector velocity-span curve, photometric curve. Values measured on the whole spectrum. Part of the spectrum used: orders $\#10$ to $\#58$. In this case with a low $v\sin{i}$ value, the bisectors are only shifted, the RV curve mimics a companion, and the photometric variation is small. A Keplerian model with 0.1\,M$_{\rm Jup}$ and a 4.4-day period on a circular orbit would fit the RV curve.}
    \label{G2_i10_t60_f1p_vsini2km}
\end{figure}

Figure~\ref{G2_i30_t30_f1p_vsini7km} (G2V-type star, $i$ = 30\degr, $\theta$ = 30\degr, 1.02\% spot, $v\sin{i}$ = 7\,km\,s$^{\rm -1}$) and Fig.~\ref{G2_i10_t60_f1p_vsini2km} (G2V-type star, $i$ = 10\degr, $\theta$ = 60\degr, 1.03\% spot, $v\sin{i}$ = 2\,km\,s$^{\rm -1}$) show examples of results obtained when the whole spectral range is used (orders $\#10$ to $\#58$). Table~\ref{G2_i90_t10to90} gives examples of values obtained for $A$ and $S$ using the whole spectra and a spot with $f_r \simeq$ 1\%, and the diagnostics that can be used to detect if it is activity-induced RV. $A/S$ is independent of $f_r$, thus those values can be scaled down. We note that the features presented in Fig.~\ref{G2_i10_t60_f1p_vsini2km} are identical to those expected from an orbiting planet (0.1\,M$_{\rm Jup}$ with a 4.4-day period). Orbiting planets produce periodic RV curves. The CCF shape does not change with time; instead, the CCFs are just shifted one to the other, consistently with a Doppler-shift of the spectra induced by the presence of a planet. 

In fact, for stars with non-resolved lines (low $v\sin{i}$), when relatively low-amplitude periodic RV variations with short periods are observed, the bisectors are just shifted and similar to those produced in the case of a planet. Additional observables are mandatory for attributing those variations either to planets or to spots.

\subsection{Discussion of various cases}

\begin{figure}[t!]
    \centering
    \includegraphics[width=1\hsize]{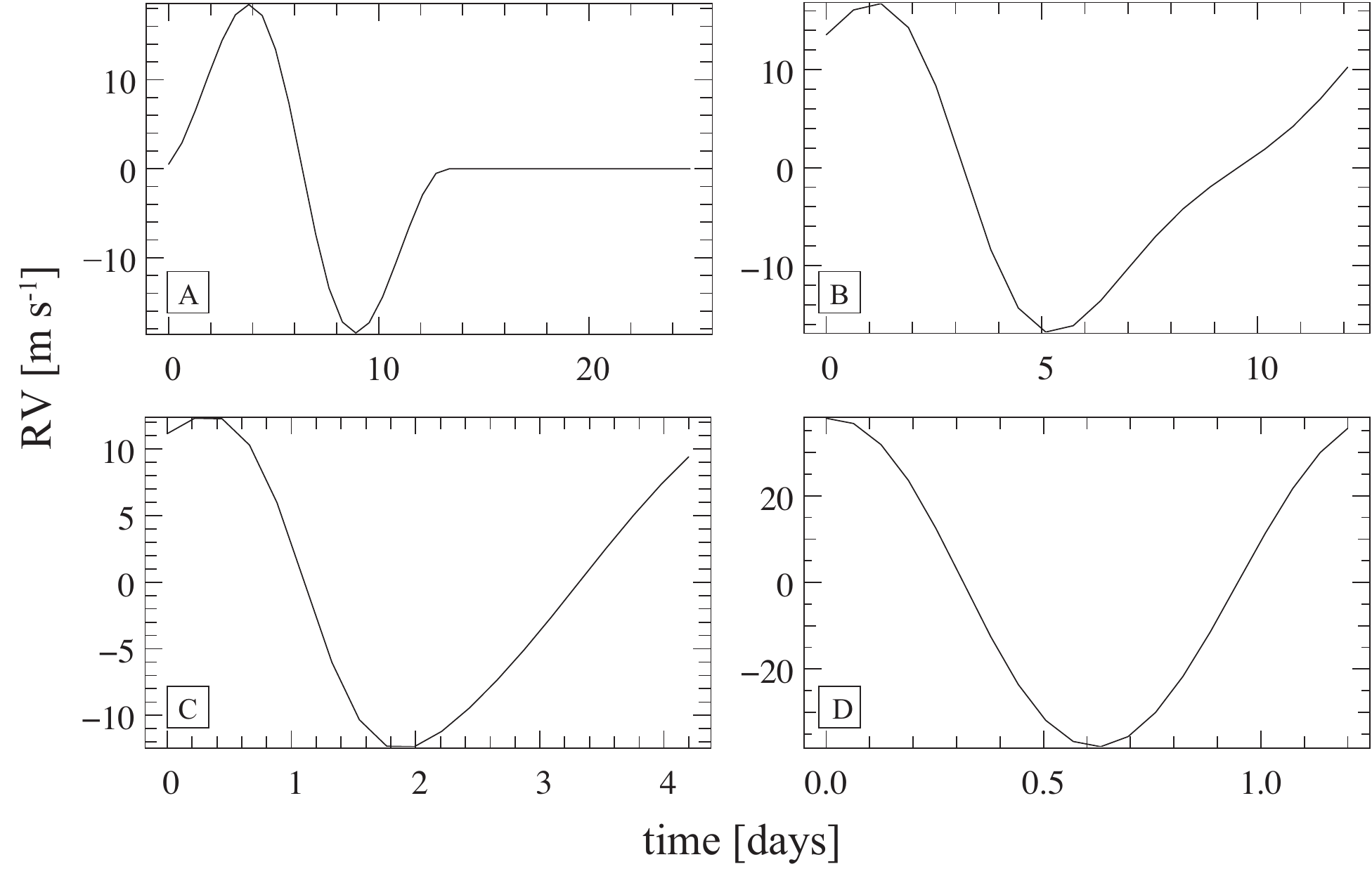}
    \caption{RV-curve shapes for various configurations (see text).}
    \label{rv_shapes}
\end{figure}

In the case of low $v\sin{i}$, for a given star inclination, the detailed shape of the RV curve depends on the spot latitude. Figure~\ref{rv_shapes} shows different shapes that can be obtained with $v\sin{i}$ = 2\,km\,s$^{\rm -1}$. We can distinguish several cases:

\begin{itemize}
\item[-] A: star at high inclination ($i$ = 90\degr). The spot is hidden part of the time, its projected size varies considerably along the rotational phase (equal to 0 when the spot is not visible), thus photometric variation is high (Table~\ref{G2_i90_t10to90}).

\item[-] B: star fairly inclined ($i$ = 30\degr). The spot is seen along the whole rotational phase (with $\theta$ = 40\degr), and its projected size does not change as much as in the previous case. Photometric variation is smaller (Table~\ref{G2_i90_t10to90}).

\item[-] C: star very inclined ($i$ = 10\degr). The spot is not very far from the equator ($\theta$ = 60\degr), it is seen along the whole rotational phase, but its projected size variation is smaller than in the previous case. Photometric variation is much smaller (Table~\ref{G2_i90_t10to90}).

\item[-] D: star very inclined ($i$ = 10\degr). The spot is close to the pole ($\theta$ = 10\degr), it is seen along the whole rotational phase, and its projected size varies very little along the rotational phase. Photometric variations are very small (Table~\ref{G2_i90_t10to90}).
\end{itemize}

\begin{table}[ht!]
    \caption{Values obtained for $A$ (m\,s$^{\rm -1}$), $S$ (m\,s$^{\rm -1}$) and $\Delta V$ (mmag) for a G2V-type star with $v\sin{i}$ = 2, 5, and 7\,km\,s$^{\rm -1}$, and diagnostics that can be used to detect if it is activity-induced RV (v: RV, in some cases; b: bisectors; p: photometry).}
    \label{G2_i90_t10to90}
    \begin{center}
\begin{tabular}{l l l l l l l l l l l}
        \hline
	\hline
        	$i$		& $\theta$		& $A$ / $S$		& $A$ / $S$		& $A$ / $S$		& $\Delta V$	\\
			&			& 2\,km\,s$^{\rm -1}$	& 5\,km\,s$^{\rm -1}$	& 7\,km\,s$^{\rm -1}$	&			\\
         (\degr)	& (\degr)	& (m\,s$^{\rm -1}$)	& (m\,s$^{\rm -1}$)	& (m\,s$^{\rm -1}$)	&			\\
        \hline
         90		& 90			& 37 / 2.1		& 103 / 43		& 155 / 108	& 20.2	\\
         			&			& v,p			& v,b,p		& v,b,p		&		\\
         60		& 60			& 34 / 1.8		& 98 / 39		& 147 / 105	& 19.5	\\
         			&			& p			& b,p			& b,p			&		\\
         30		& 40			& 34 / 2		& 97 / 20		& 145 / 100	& 14		\\
         			&			& p			& b,p			& b,p			&		\\
         30		& 30			& 31 / 1.9		& 92 / 39		& 145 / 104	& 13		\\
         			&			& p			& b,p			& b,p			&		\\
         10		& 60			& 25 / 1.1		& 65 / 21		& 94 / 51		& 5.7		\\
         			&			& p			& b,p			& b,p			&		\\
	10		& 10			& 16 / 1.2		& 48 / 27		& 76 / 75		& 1.8		\\
         			&			& p?			& b,p?		& b,p?		&		\\
        \hline
      \end{tabular}
    \end{center}
\end{table}

Apart from the first curve (A), we can find a Keplerian model that is acceptable, even if those with small $i$ or  with a spot near the pole are better. For intermediate inclinations, the departure from a 1-planet Keplerian model is small and the Keplerian fits are quite acceptable (Fig.~\ref{G2_i30_t30_f1p_vsini7km}).

Concerning the RV and bisector velocity-span amplitudes, it can be seen that, for a given $v\sin{i}$, $A$ and $S$ decrease when the star and spot inclinations decrease. Moreover, $A$ and $S$ have different values depending on $v\sin{i}$.
More precisely, we can distinguish several cases\footnote{We assume that $A$ or $S$ will be detectable with good confidence if they are larger than 6\,m\,s$^{\rm -1}$ (with {\small HARPS}); we recall that $A$ and $S$ refer to peak-to-peak amplitudes}:

\begin{itemize}
\item[-] For large $v\sin{i}$ (around 7\,km\,s$^{\rm -1}$), $A/S\sim 1.0-1.5$; hence, when RV variations are detected, bisector velocity-span variations will also be detected in most cases, but for very small spots where it is below the detection limit while $A$ is still detected. At high inclination, we may find cases where variations in $A$ are detected, whereas those of $S$ are not ({\it e.g.}, $f_r$ = 0.1\%, $i$ = 90\degr). However in such cases, variations will not be observed over the whole phase, so there is no risk of confusion between planets and spots, except in very particular and unlikely configurations with several spots. At very low inclinations ($i \sim$ 10-20\degr) and low $\theta$, $A/S\sim 1$, which means that, if RV variations are detected, variations in $S$ with similar amplitudes can be detected.

\item[-] For low $v\sin{i}$ (2\,km\,s$^{\rm -1}$), $A/S\gg 5$, and it is independent of $f_r$; $S$ is always smaller than 3\,m\,s$^{\rm -1}$, whereas $A$ is larger than 6\,m\,s$^{\rm -1}$ for $f_r$ = 1\%. Moreover, in the case of low inclinations of star and spot, the photometric variations may become smaller than 5\,mmag, {\it {\it i.e.}}, difficult to detect routinely. In the example shown in Fig.~\ref{G2_i10_t60_f1p_vsini2km} ($i$ = 10\degr, $\theta$ = 60\degr), the amplitude of photometric variations is 5.7\,mmag (peak to peak), twice as small a spot would produce twice as small a photometric variation.

\item[-] For intermediate $v\sin{i}$ (5\,km\,s$^{\rm -1}$), $A$ and $S$ variations are always detectable for $f_r$ = 1\%. A simple scaling shows, however, that for much lower values of $f_r$ ({\it e.g.}, 0.2\%), RV variations could be detectable while $S$ would not. This would happen for low values of $i$ and $\theta$, and the amplitude of photometric variations would become undetectable (at a 0.4\,mmag level for a 0.2\% spot). It is thus mandatory to carry out photometric measurements below the 1\,mmag level to detect such problematic cases.

\item[-] There still remains a narrow range of situations where $A$ can be detected, whereas $S$ cannot. This would happen for $i \sim$ 30\degr, $f_r \sim$ 0.02\%. In that case $A\sim 3\,{\rm m}\,{\rm s}^{\rm -1}$ and $S\sim 2\,{\rm m}\,{\rm s}^{\rm -1}$. And the corresponding $\Delta V$ would be on the order of 0.3\,mmag, hence undetectable.
\end{itemize}

\section{Equatorial spot on other types of stars}

We computed various simulations for two other types of stars, namely a F6V and a K2V. Here we present the results of those simulations and analyze them.

\subsection{F6V-type star}

We assumed an edge-on F6V-type star with a 1\% spot. The star's projected rotational velocity was assumed to be 2, 7, or 20\,km\,s$^{\rm -1}$. We assumed as before a difference of 1200\,K between the star and the spot temperatures: $T_{\rm eff}$ = 6200\,K, $T_{\rm spot}$ = 5000\,K. We also assumed $v_{\rm macro}$ = 1.5\,km\,s$^{\rm -1}$.

\begin{figure}[t!]
    \centering
    \includegraphics[width=0.9\hsize]{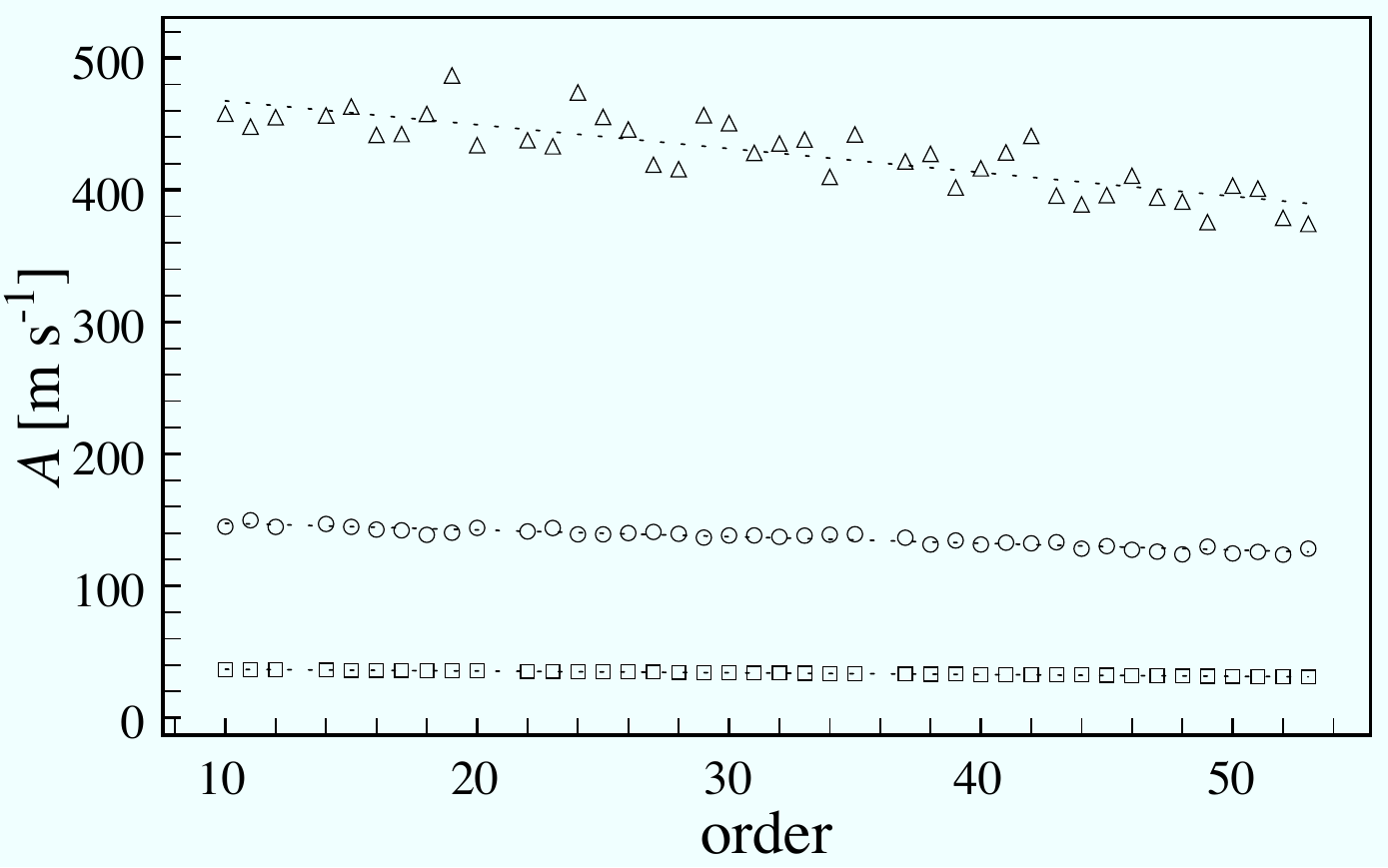}
    \includegraphics[width=0.9\hsize]{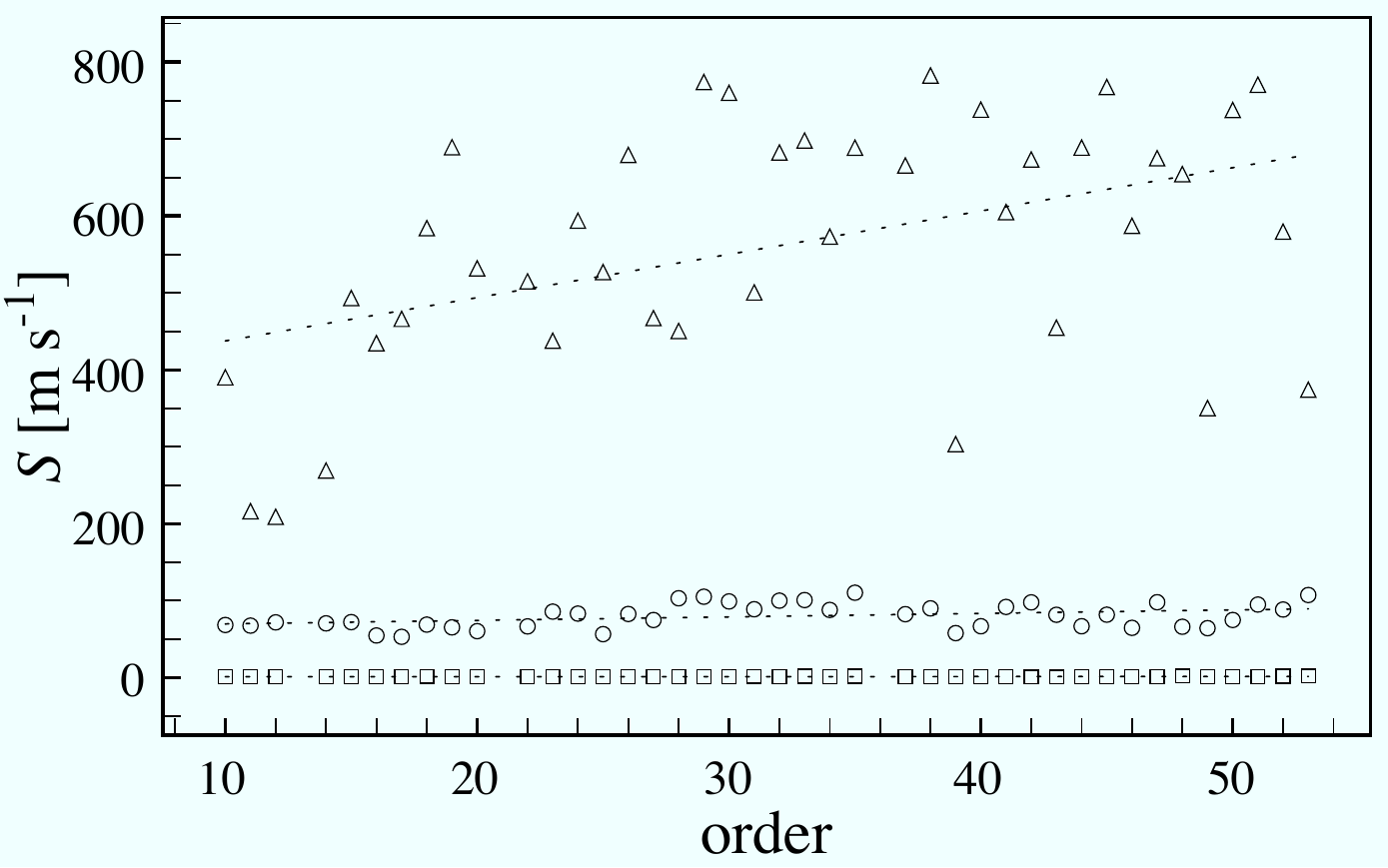}
    \caption{Measured $A$ and $S$ as a function of order number; assuming a 1\% equatorial spot on an edge-on F6V-type star with $v\sin{i}$ = 2 (squares), 7 (circles), and 20\,km\,s$^{\rm -1}$ (triangles).}
    \label{AS_F6_90}
  \end{figure}

Figure~\ref{AS_F6_90} shows the values of $A$ and $S$ obtained for different $v\sin{i}$. We find the following relations between $A$, $S$, $f_r$, and $v\sin{i}$, with $v_0$ still fixed to 3\,km\,s$^{\rm -1}$ (see Fig.~\ref{AS-vsini_F6_G2_K2}):

\begin{equation}
\label{A_F6}
$$A = 15.4 \, f_r \, (v\sin{i})^{1.1}$$,
\end{equation}

\begin{equation}
\label{S_F6}
$$S = 7.1 \, f_r \, \left(\sqrt{(v\sin{i})^2+v_0^2} - v_0 \right)^{1.5}$$.
\end{equation}

\begin{figure}[t!]
    \centering
    \includegraphics[width=1\hsize]{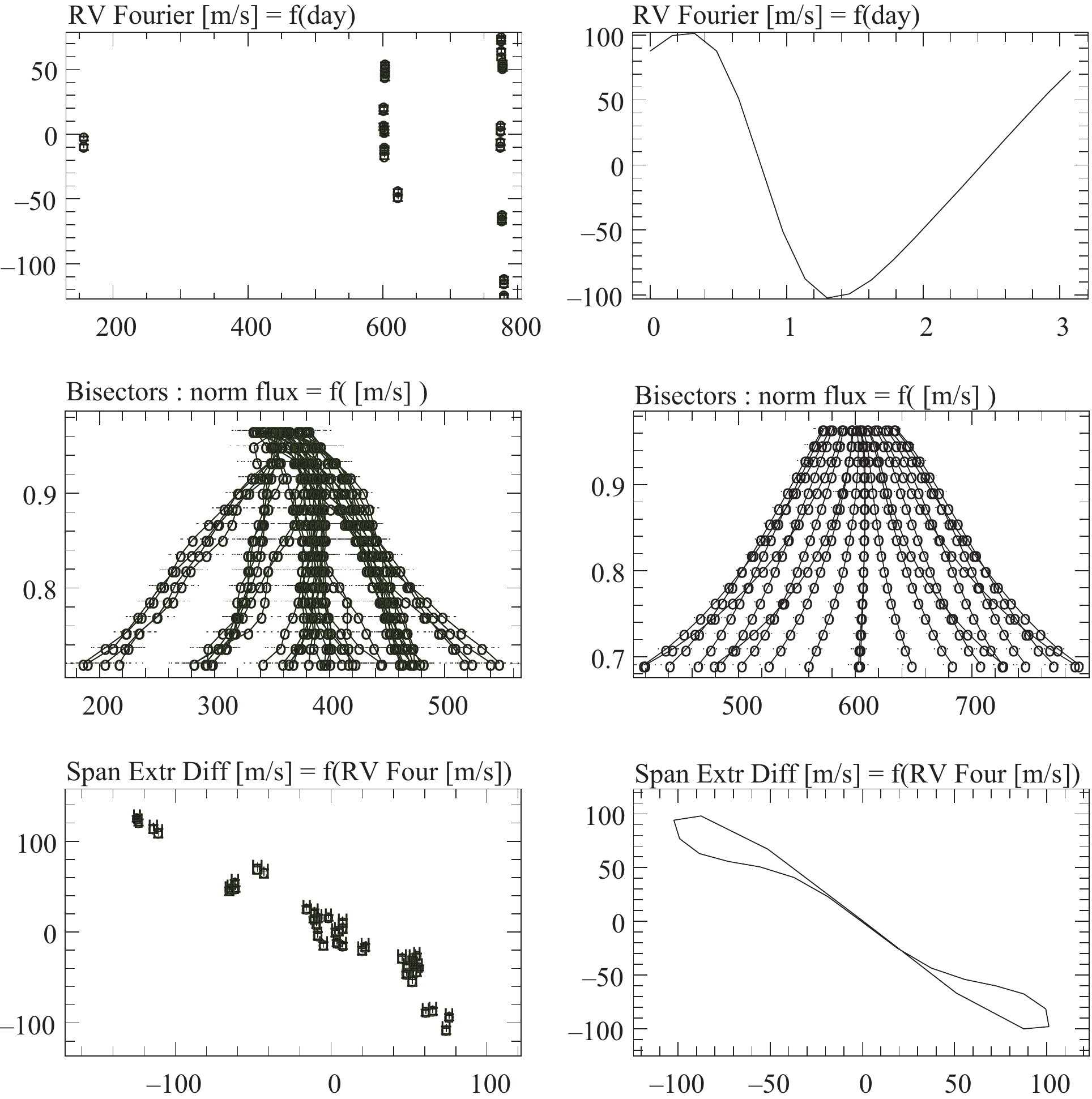}
    \caption{Example of comparison of an active F7V-type star ($v\sin{i} \sim$ 9.6\,km\,s$^{\rm -1}$) between real observations (left) and the present simulations (right): we obtain similar RV variation amplitudes (top), CCF behavior (center) and correlation between bisector velocity-span and RV (bottom).}
    \label{AS_F6-real}
  \end{figure}

In Fig.~\ref{AS_F6-real}, we show an example of comparing with real observations and the present simulations: the RV of an F7V-type star with $v\sin{i}$ = 9.6\,km\,s$^{\rm -1}$,\footnote{Estimated from the CCF, \cite{glebocki00} (2000) gives $v\sin{i}$ = 7\,km\,s$^{\rm -1}$, the simulation uses 11\,km\,s$^{\rm -1}$ to reproduce the slope of the correlation.} measured with {\small HARPS} and our software, are variable (left, top), but the profile of the CCFs changes as a function of time (left, bottom). Our simulations of the same kind of star show that these variations (right, top: phased) can result from the presence of a spot induced by stellar activity: we obtain the same behavior of the CCF and the same correlation between the corresponding bisector velocity-span and RV (right, bottom), with the same amplitude levels. The value of $\log R_{HK}'$ is rather high, namely $-4.3$, which effectively indicates a relatively high level of activity. We note that this is coherent with the high level of amplitude variations ($>$ 200\,m\,s$^{\rm -1}$ peak to peak).

\subsection{K2V-type star}

\begin{figure}[t!]
    \centering
    \includegraphics[width=0.9\hsize]{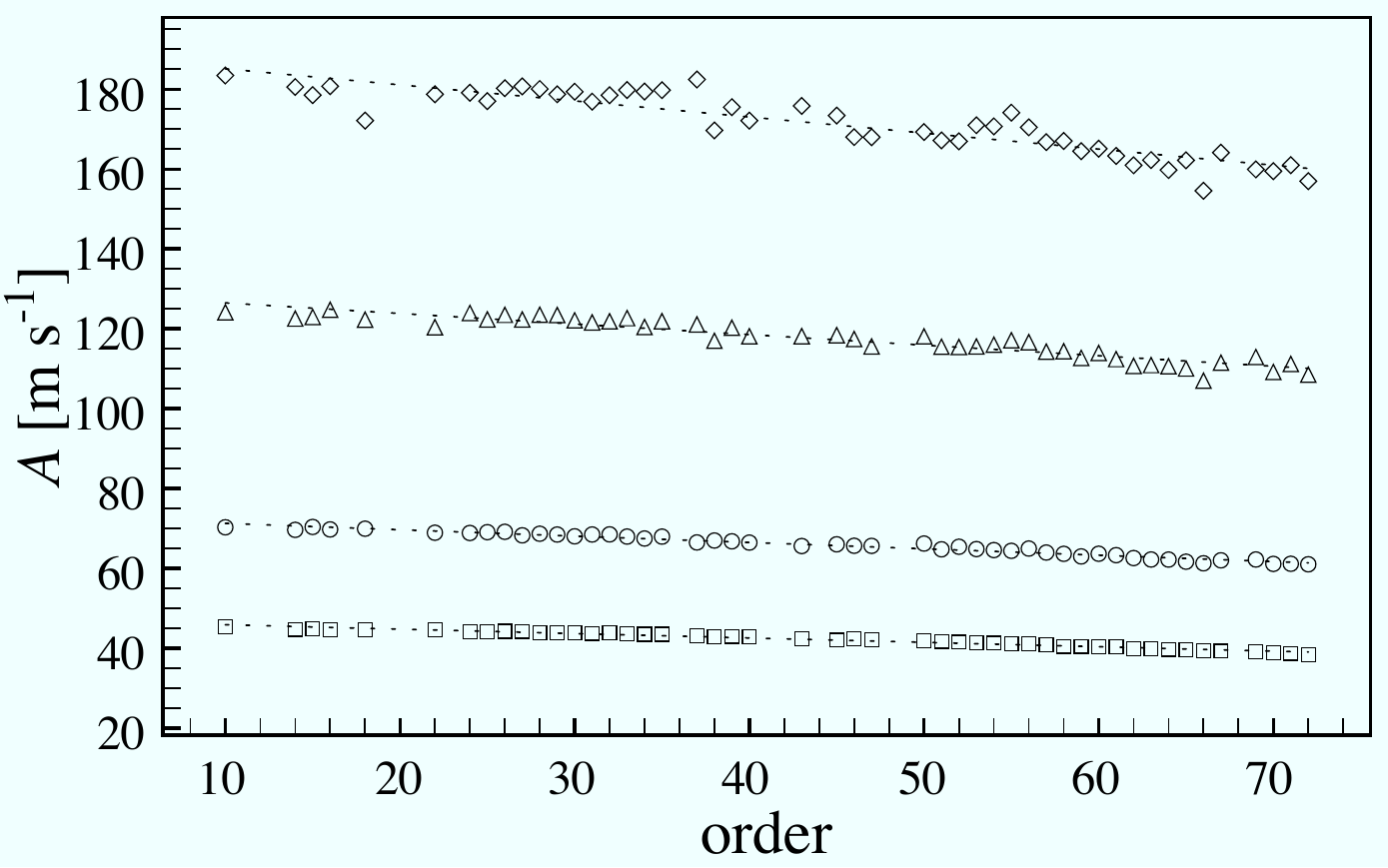}
    \includegraphics[width=0.9\hsize]{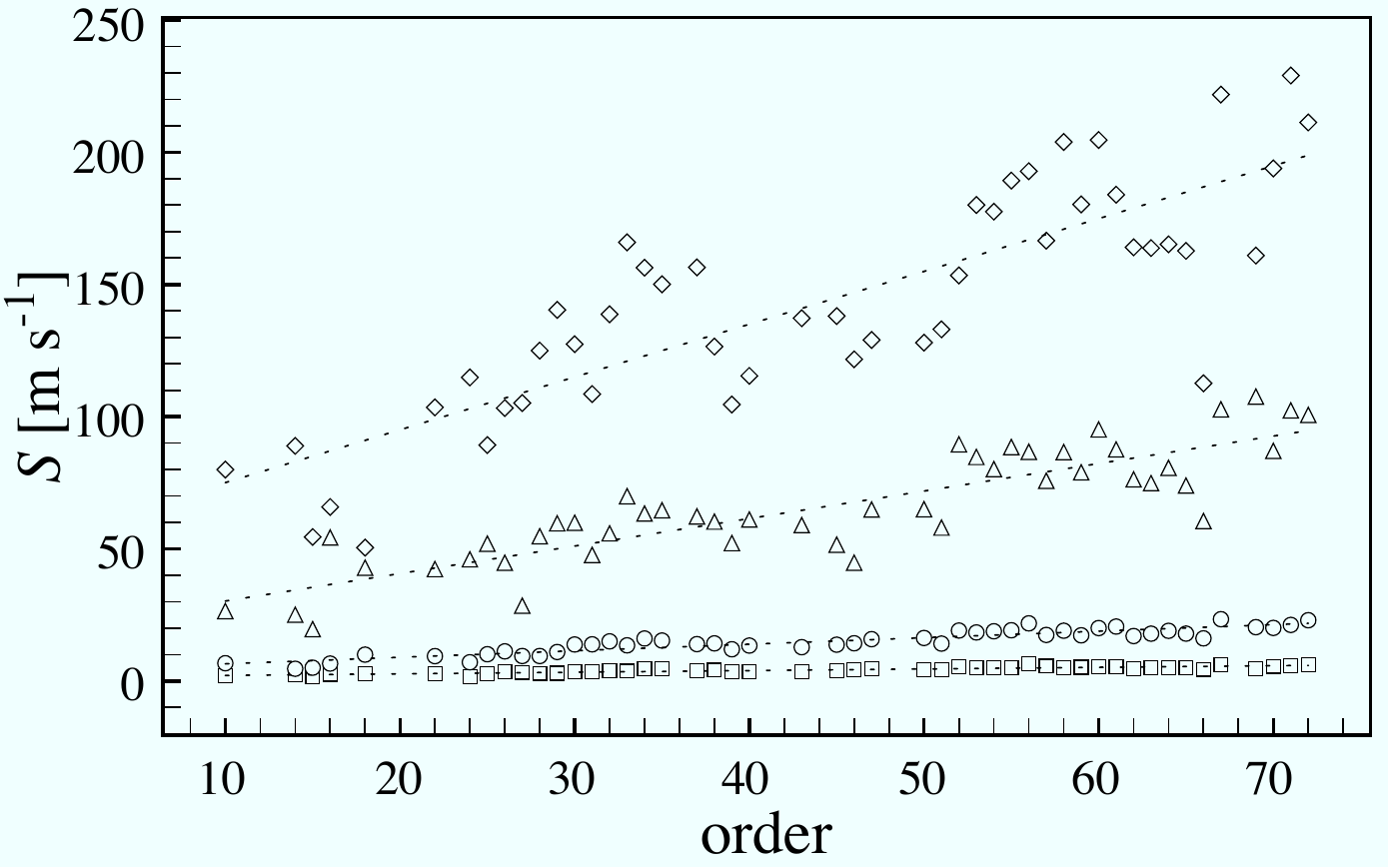}
    \caption{Measured $A$ and $S$ as a function of order number; assuming a 1\% equatorial spot on an edge-on K2V-type star with $v\sin{i}$ = 2 (squares), 3 (circles), 5 (triangles), and 7\,km\,s$^{\rm -1}$ (diamonds).}
    \label{AS_K2_90}
  \end{figure}

\begin{figure}[t!]
   \centering
   \includegraphics[width=0.7\hsize]{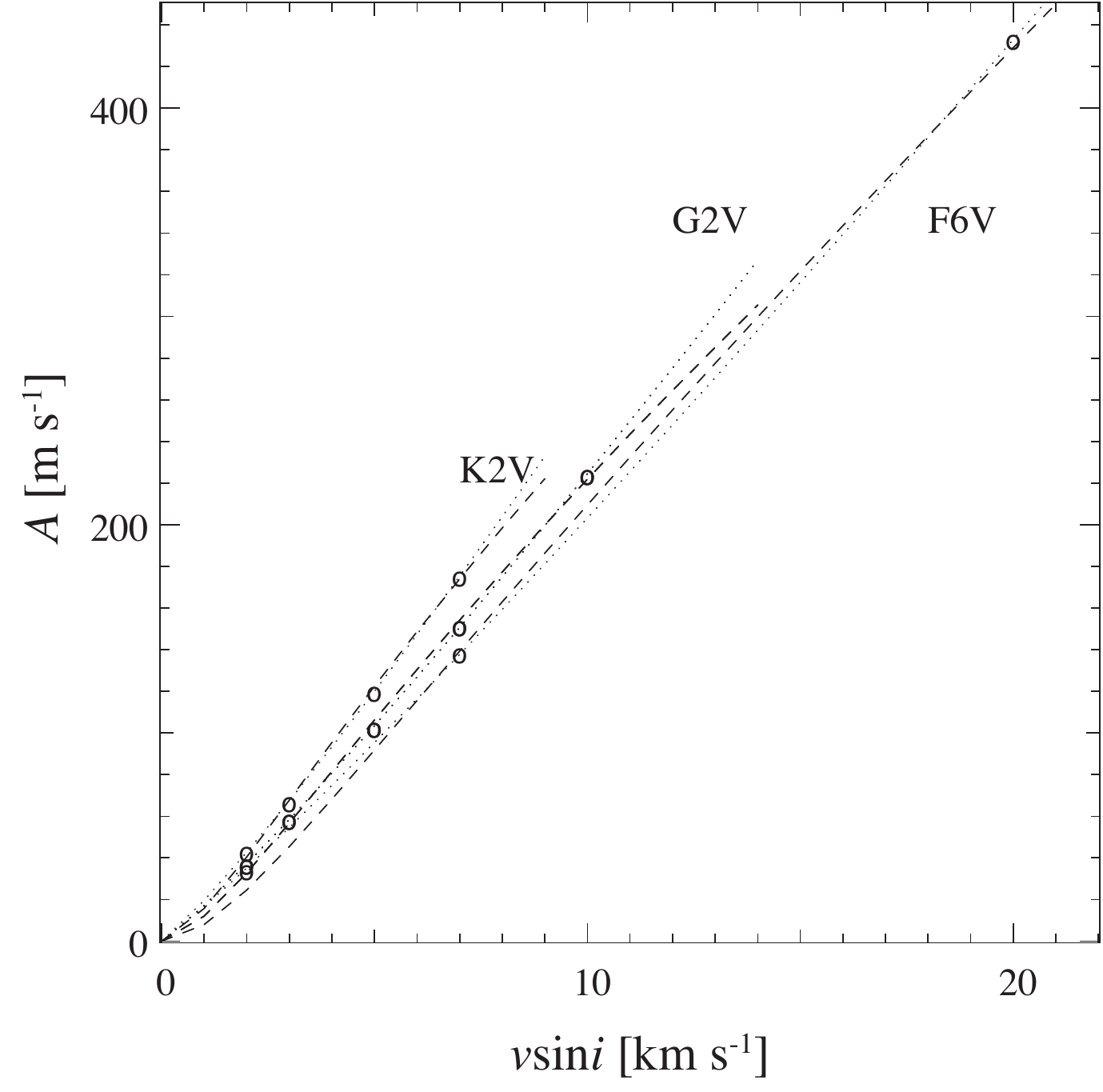}
   \includegraphics[width=0.7\hsize]{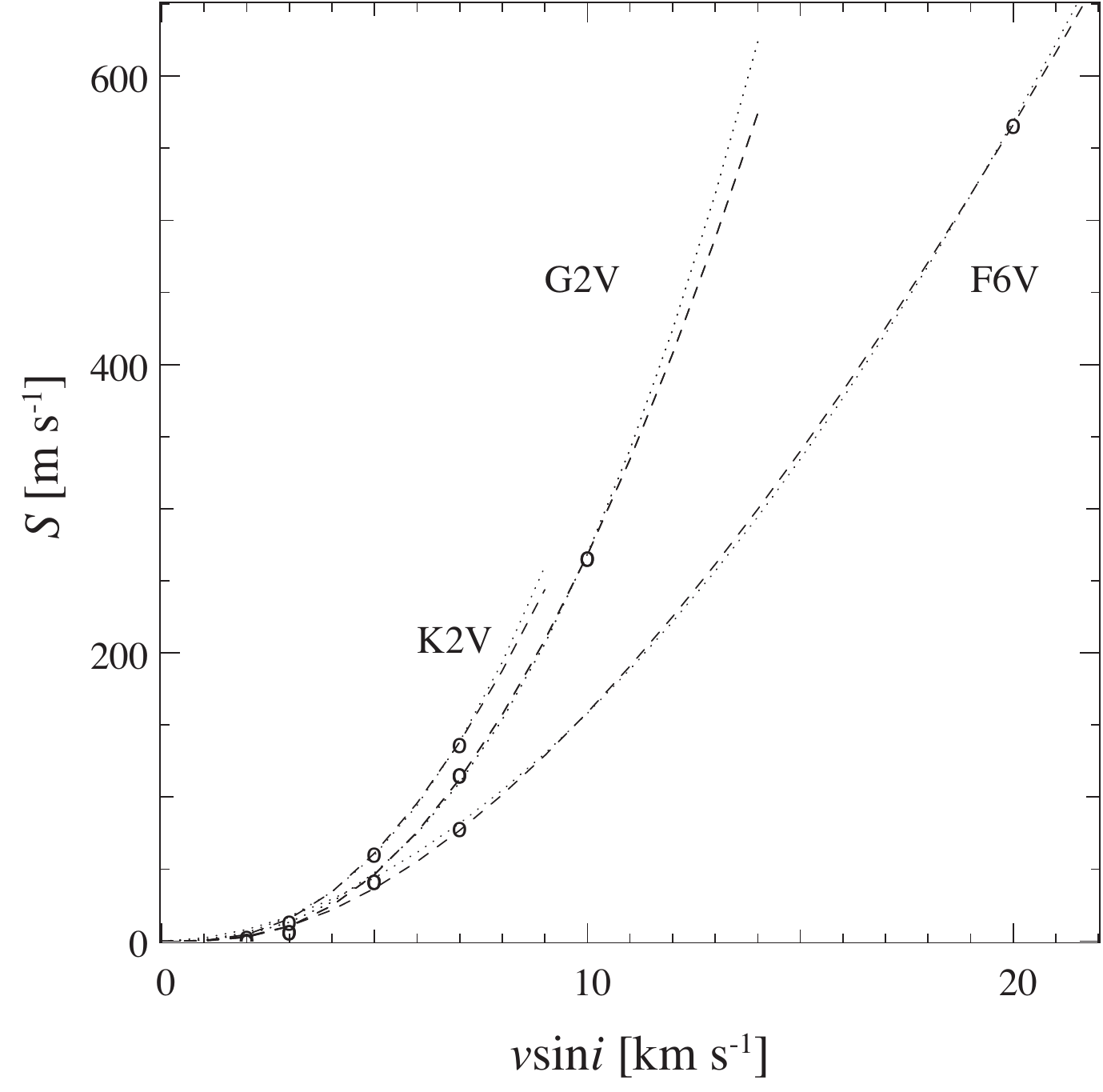}
\caption{$A$ and $S$ as a function of the star's projected rotational velocity, for an equatorial spot on edge-on F6V, G2V, and K2V-type stars. $A$ varies approximately linearly with $v\sin{i}$. $S$ varies as $\left(\sqrt{(v\sin{i})^2+v_0^2} - v_0 \right)^{\alpha}$.
If we take into account the instrumental resolution (dashed lines) ${\alpha} \simeq$ 1.5 for the F6V-type star, 1.7 for the K2V, otherwise (dotted lines) ${\alpha} \simeq$ 2.4, and 2.5. Values measured on orders $\#10$ to $\#58$.}
    \label{AS-vsini_F6_G2_K2}
  \end{figure}

We also conducted full-order simulations in the case of a K2V-type star with $T_{\rm eff}$ = 4800\,K, $T_{\rm spot}$ = 3600\,K, and $v_{\rm macro}$ = 0.9\,km\,s$^{\rm -1}$. The star was assumed to be seen edge-on and the spot to be equatorial. Different values of $v\sin{i}$ were used: 2, 3, 5, and 7\,km\,s$^{\rm -1}$. Figure~\ref{AS_K2_90} shows the values of $A$ and $S$ obtained for different $v\sin{i}$. We find the following relations between $A$, $S$, $f_r$, and $v\sin{i}$, with $v_0$ still fixed to 3\,km\,s$^{\rm -1}$ (see Fig.~\ref{AS-vsini_F6_G2_K2}):

\begin{equation}
\label{A_K2}
$$A = 18.3 \, f_r \, (v\sin{i})^{1.1}$$,
\end{equation}

\begin{equation}
\label{S_K2}
$$S = 10.0 \, f_r \, \left(\sqrt{(v\sin{i})^2+v_0^2} - v_0 \right)^{1.7}$$.
\end{equation}

\begin{figure}[t!]
    \centering
    \includegraphics[width=1\hsize]{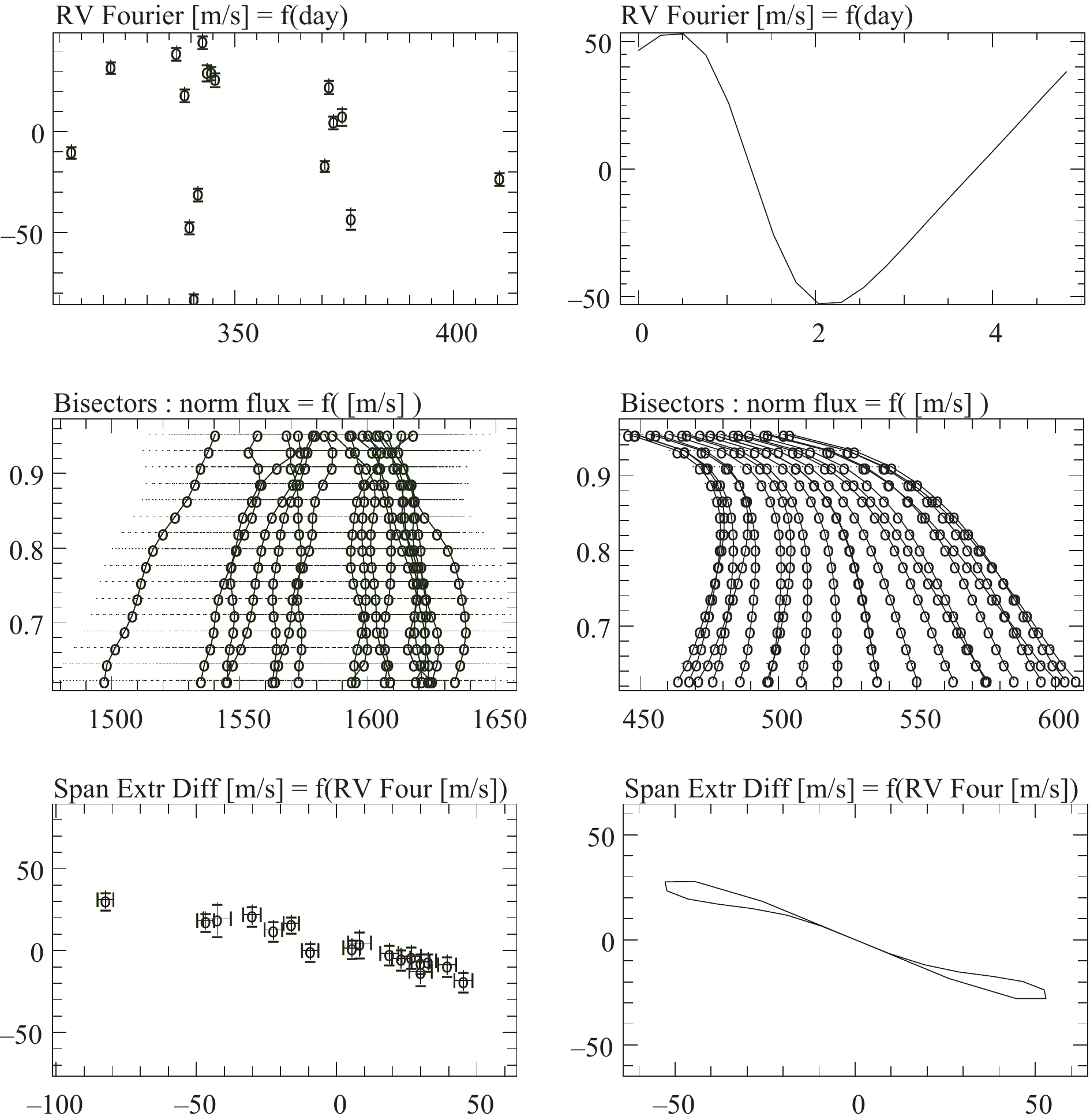}
    \caption{
    Example of comparison of an active K1V-type star ($v\sin{i} \sim$ 4.8\,km\,s$^{\rm -1}$) with real observations (left) and the present simulations (right): we obtain similar RV variation amplitudes (top), CCF behavior (center), and correlation between the bisector velocity-span and RV (bottom).}
    \label{AS_K1-real}
\end{figure}

As an illustration, we tried to find whether we could explain the RV and bisector velocity-span variations observed with {\small HARPS} in the case of a K1V-type star. Figure~\ref{AS_K1-real} shows such an example: the RV, measured with {\small HARPS} and our software, are variable (top) with an amplitude of 130\,m\,s$^{\rm -1}$, but the profile of the CCF changes as a function of time (center and bottom), with a bisector velocity-span amplitude of 50\,m\,s$^{\rm -1}$. We find that a close-to-equatorial, 1\% sized spot located on a star rotating with $v\sin{i}$ = 5\,km\,s$^{\rm -1}$ could produce similar amplitudes. Note that the value of $\log R_{HK}'$ is $-4.3$ for this star, which is consistent with the presence of a relatively high level of activity.

\subsection{Comparison to G2V-type stars}

\begin{table}[t!]
    \caption{Comparison between F, G and K-type stars.}
    \label{F-G-K_i90t90f1}
    \begin{center}
\begin{tabular}{l l l l}
        \hline
	\hline
						& F6V	& G2V	& K2V	\\
						& (m\,s$^{\rm -1}$)	& (m\,s$^{\rm -1}$) & (m\,s$^{\rm -1}$)\\
	\hline
	$A$ (2\,km\,s$^{\rm -1}$)	& 34		& 37		& 43		\\
	$S$					& 1.1		& 2.1		& 3.9		\\
	$A$ (3\,km\,s$^{\rm -1}$)	& \dots	& 59		& 67		\\
	$S$					& \dots	& 7.9		& 14		\\
	$A$ (5\,km\,s$^{\rm -1}$)	& \dots	& 103	& 120	\\
	$S$					& \dots	& 43		& 62		\\
	$A$ (7\,km\,s$^{\rm -1}$)	& 138	& 155	& 175	\\
	$S$					& 79		& 108	& 138	\\
	$A$ (10\,km\,s$^{\rm -1}$)	& \dots	& 224	& \dots	\\
	$S$					& \dots	& 267	& \dots	\\
	$A$ (20\,km\,s$^{\rm -1}$)	& 433	& \dots	& \dots	\\
	$S$					& 567	& \dots	& \dots	\\
	$\Delta V$ (mmag)		& 19.1	& 20.2	& 25.1	\\
	\hline
      \end{tabular}
    \end{center}
\end{table}

\begin{figure}[t!]
    \centering
    \includegraphics[width=0.9\hsize]{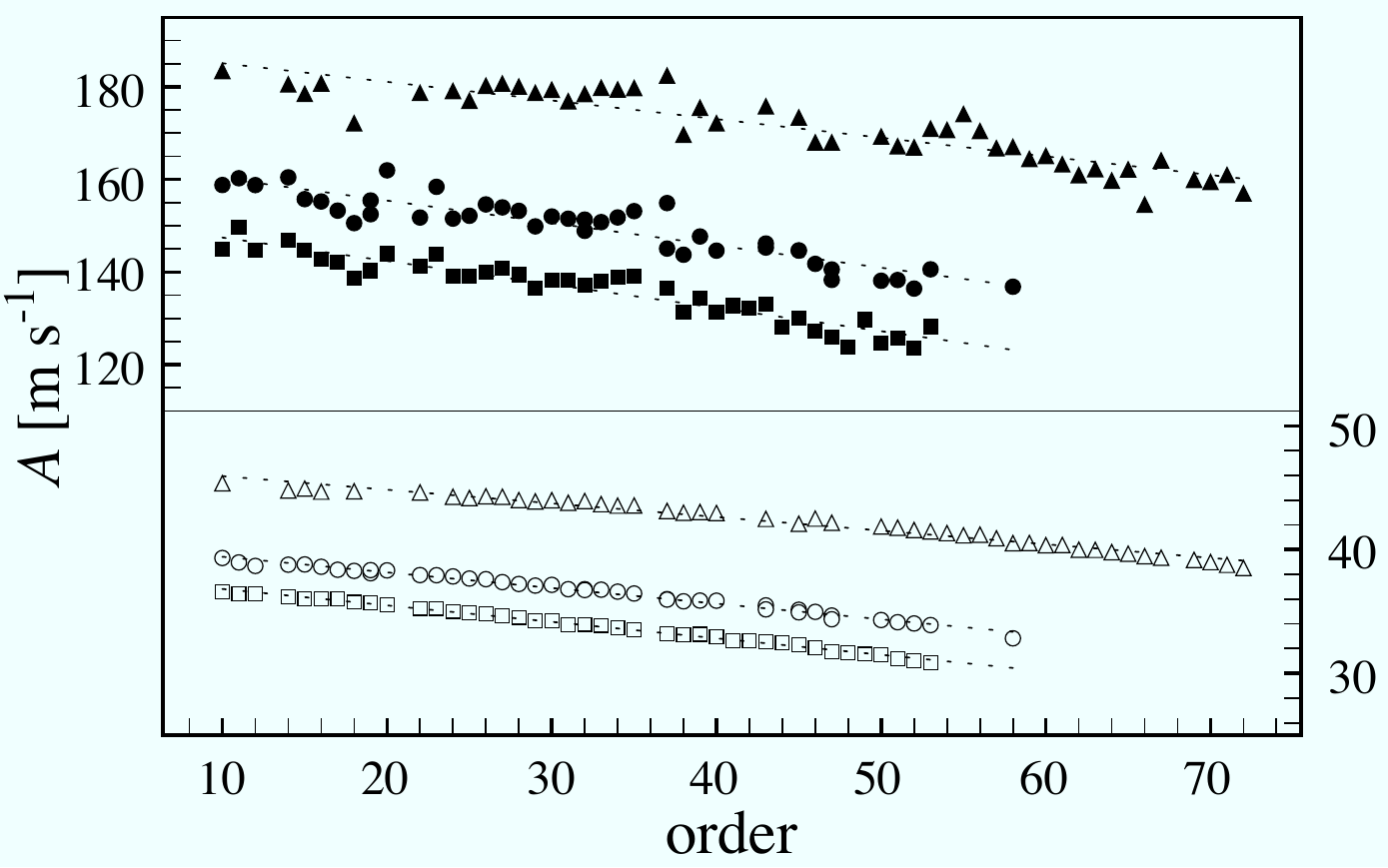}
    \includegraphics[width=0.9\hsize]{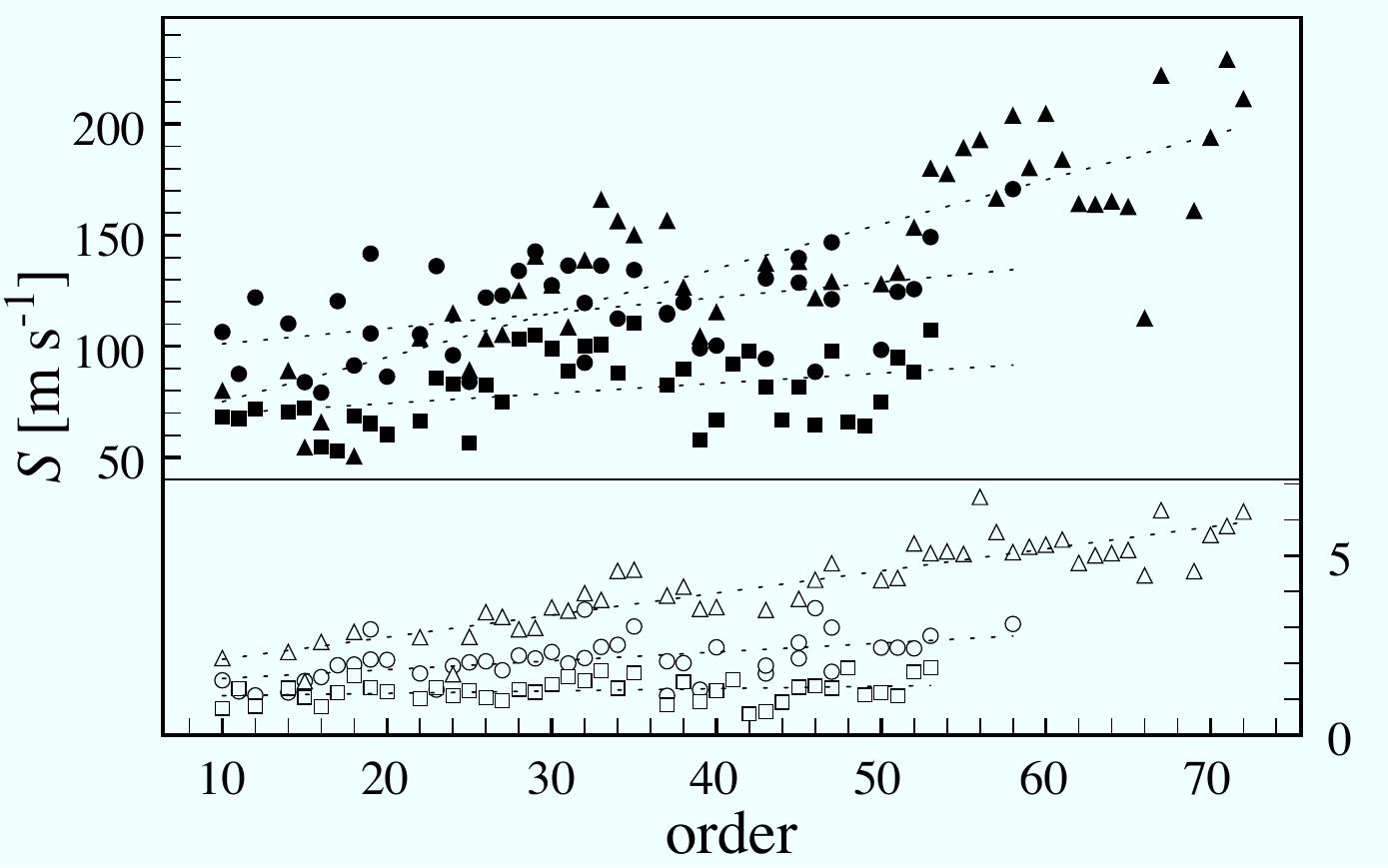}
    \caption{Comparison between F6V (squares), K2V (triangles), and G2V (circles)-type stars for two $v\sin{i}$ (open symbols: 2\,km\,s$^{\rm -1}$, filled symbols: 7\,km\,s$^{\rm -1}$). Orders $\#10$ to $\#72$.}
    \label{AS_F6-K2-G2_90}
\end{figure}

Table~\ref{F-G-K_i90t90f1} summarizes the values of $A$, $S$, and $\Delta V$ obtained for these edge-on F6V and K2V-type stars with various $v\sin{i}$ and a 1\% equatorial spot. It allows for comparison with G2V-type star (second column). Figure~\ref{AS_F6-K2-G2_90} shows the comparison between F6V and K2V spectral types with respect to G2V order-to-order $A$ and $S$ measurements.
Globally, one sees that, for a given set of conditions (spot location and size, star inclination, and $v\sin{i}$, $T_{\rm eff}$, and $T_{\rm spot}$), the amplitude of RV and bisector velocity-span variations increase when we go from F6V to K2V-type stars. For a given $v\sin{i}$, this means that it will be more difficult to detect bisector velocity-span variations in the case of F6V-type stars than in the case of G2V-type stars and easier for K2V-type stars. However, one has to note that F6V-type stars rotate an average faster than G2V-type stars. The occurrence of cases where possible confusions between planets and spots should thus decrease, whereas K2V-type stars rotate an average slower than G2V-type stars, so the occurrence of those cases should increase.

\section{Impact on the search for short-period RV planets and further studies}

In the previous sections, we have shown that spots can produce a variety of features (RV, bisector shapes, and variations), whose characteristics vary according to the spots and star characteristics.
We have also seen that the precise values of RV and, to a greater extent, the values of the bisector velocity-span depend on the line or set of lines taken into account and on the spectrograph used. A quantitative comparison of the bisector velocity-span or bisector velocity-span to RV correlation should be made for the same line or set of lines, acquired with the same spectrograph (or at least with the same spectral resolution) and analyzed with the same software. Working on the full CCF allows individual effects to be averaged out and allows quantitative comparison to models, provided the models assume spectral types and projected rotational velocities identical to those of the star under study and use the same spectral lines and same spectral resolution.

From our studies, spots with typical sizes (1\%) at the surface of stars with high $v\sin{i}$ will be easily identified and may be characterized using the criteria presented above (bisectors, bisector velocity-spans). In the case of stars with low inclinations and spots near the pole, confusion may however arise when the RV amplitudes are small ($<$ 10\,m\,s$^{\rm -1}$). The situation is more complex in the case of stars with low $v\sin{i}$.

For stars with low $v\sin{i}$, spot features may, in some cases, mimic those produced by planets. This happens when 1) the observed RV variations have periods similar to the star rotational period, 2) variations are observed in the RV curves, and at the same time 3) bisectors do not change in shape, but are just shifted according to the RV changes. From the study above, this happens in the case of stars with intermediate $v\sin{i}$ (typ. 5\,km\,s$^{\rm -1}$) and a very small (0.1\%) spot or in the case of stars with low $v\sin{i}$ compared to the spectrograph resolution (typ. $\leq$ 3\,km\,s$^{\rm -1}$ for a resolution of 100\,000, $\leq$ 6\,km\,s$^{\rm -1}$ for a resolution of 50\,000), even with a spot that has a common (typ. $\leq$ 1\%) size. When the spot is large enough and its location favorable, photometry will in general tell whether the variations are due to spots or not. In the case of small spots and for particular inclinations of the star, the photometric precision may be out of reach.
Of course, the occurrence of such cases will increase when searching for planets with lower and lower masses (super-Earths or Earths) for a given orbit, as they will produce smaller RV amplitudes and bisector shapes comparable to those of smaller spots, {\it i.e.}, with small $S$.
Note that indicators of activity such as $\log R_{HK}'$ may not be sensitive enough for low levels of activity (hence the low amplitude of RV variations). Indeed, for levels of activity as low as $\log R_{HK}' \sim -4.8$, the level of the amplitude for the RV variations is still a few m\,s$^{\rm -1}$ (\cite{santos00} 2000, \cite{wright03} 2003).

\begin{figure}[t!]
    \centering
    \includegraphics[width=0.7\hsize]{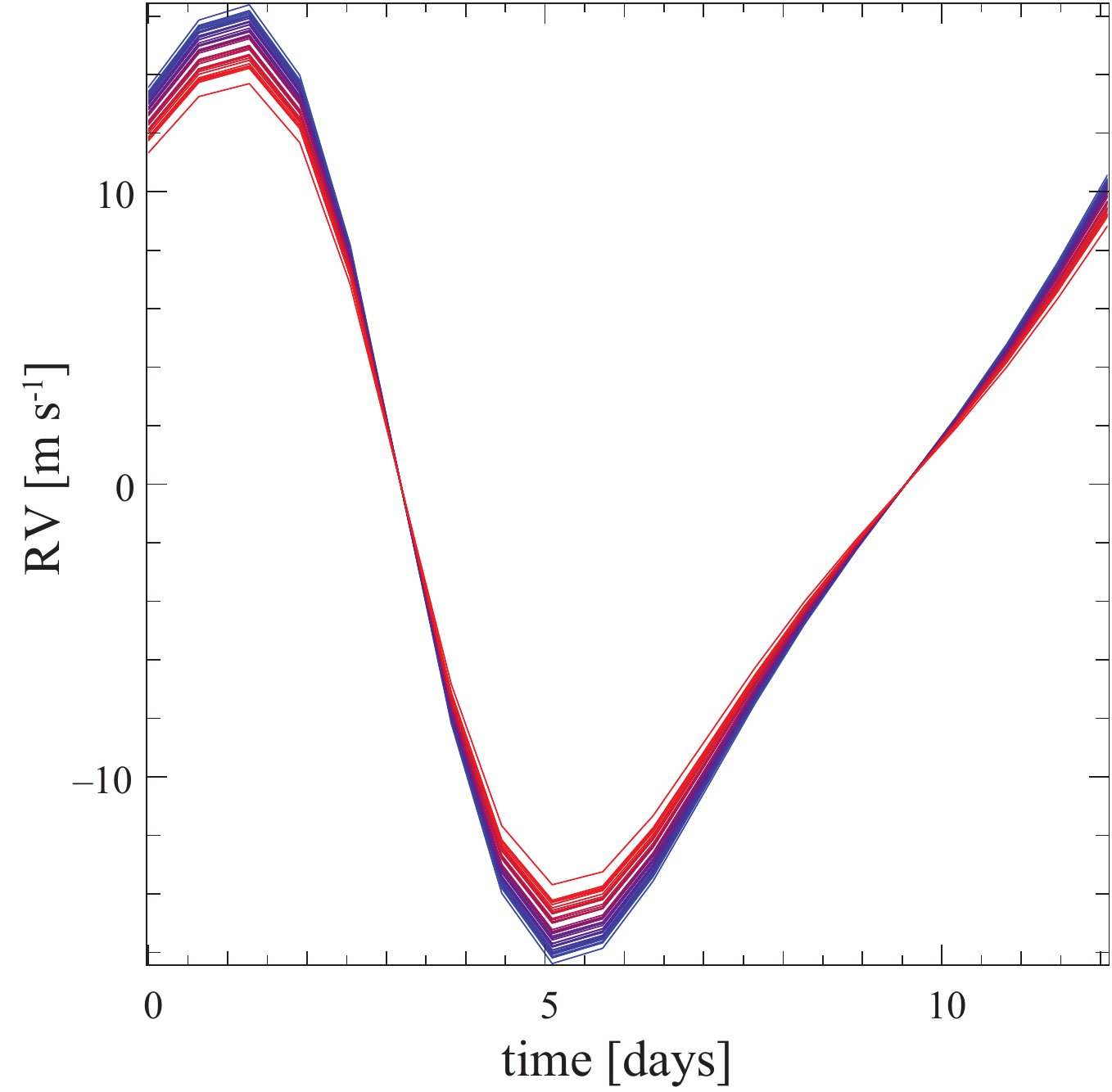}
    \caption{RV curves as a function of order number. The highest amplitude corresponds to the lowest (blue) order and the lowest amplitude to the highest (red) order. Here $\Delta V$ = 1.3\%. Hypotheses: G2V-type star, $v\sin{i}$ = 2\,km\,s$^{\rm -1}$, $i$ = 30\degr, $\theta$ = 30\degr, $f_r$ = 1\%.}
    \label{G2_i30t30f1}
  \end{figure}

The chromatic dependence of $A$ may in such cases help in distinguishing between planets and stellar spots. Indeed, in the case of a planet perturbation, no such chromatic dependence is expected, whereas in the case of spots, chromatic effects will occur. Figure~\ref{G2_i30t30f1} for example shows the RV curves as a function of order number in the case of a G2V-type star ($v\sin{i}$ = 2\,km\,s$^{\rm -1}$) seen with $i$ = 30\degr~and a 1\% spot located at $\theta$ = 30\degr. This spot would produce a photometric variation of 1.3\%. The peak-to-peak difference between the measurements of $A$ in the first 4 orders and the last 4 orders is 3.8\,m\,s$^{\rm -1}$, hence could be detectable with good signal-to-noise data. This provides a possible additional criterion for testing the origin of RV variations when bisector velocity-span and/or photometric criteria cannot be applied. It has the great advantage of being an observable present in the spectroscopic data themselves.

Will it be possible to invert an observed data set (RV variations, bisector velocity-span variations, possible photometric variations)? The unknown parameters are $i$, number of spots, temperature of spot, spot colatitude, and spot size. We already know $v\sin{i}$ increases with earlier spectral types. One spot then results in 5 free parameters and 2 spots result in 8 free parameters, so it will probably be very difficult to perform this inversion reliably. Nevertheless, our study shows that, with a given set of observations, it is possible to make the diagnostic of the presence of spots, provided we are able to measure the chromatic impact on $A$.

This study has assumed a very simple case of a single spot. Of course, the reality is more complex: the star may be covered by several spots with different temperatures, at different latitudes; they may also have inhomogeneities distributed in complex patterns; in addition the star may undergo complex patterns of pulsation. Simulations will be performed in the future to explore a wider variety of cases and to test more realistic cases.


\begin{acknowledgements}

We acknowledge support from the French CNRS and the Programme National de Plan\'etologie ({\small PNP, INSU}). This work was also funded by the French Agence Nationale pour la Recherche, ANR. These results have made use of the SIMBAD database, operated at the CDS, Strasbourg, France.
 
\end{acknowledgements}

\begin{appendix}

\section{Linear fits}

\begin{table}[h]
\caption{Values $a$ and $b$ of the linear fits $y = ax + b$ corresponding to the different plots.}
\label{linear_fits}
\begin{center}
\begin{tabular}{cc}

	\begin{tabular}[t]{l r r}
        \hline
	\hline
	Reference		& $a$	& $b$	\\
	\hline
	\multicolumn{3}{c}{Fig.~\ref{AS_G2_90}}\\
	\hline
	$A$, $\Box$		& -0.13	& 41		\\
	$A$, $\Diamond$	& -0.22	& 65		\\
	$A$, $\circ$		& -0.31	& 111	\\
	$A$, $\vartriangle$	& -0.48	& 165	\\
	\hline
	$S$, $\Box$		& 0.02	& 1.3	\\
	$S$, $\Diamond$	& 0.08	& 5.5	\\
	$S$, $\circ$		& 0.45	& 28		\\
	$S$, $\vartriangle$	& 0.69	& 94		\\
	\hline
	\multicolumn{3}{c}{Fig.~\ref{G2_i90_t90_fr1_AandS_fr}}\\
	\hline
	$A$, $\bullet$	& 141	& 0	\\
	$S$, $\circ$	& 108	& 0	\\
	\hline
	\multicolumn{3}{c}{Fig.~\ref{Results_psf_opo_AandS_N_v}, top}\\
	\hline
	$A$, $\times$	& -0.17	& 60.3	\\
	$A$, $\circ$	& -0.22	& 64.8	\\
	$A$, $\bullet$	& -0.22	& 65.5	\\
	\hline
	$S$, $\times$	& 0.06	& 2.6	\\
	$S$, $\circ$	& 0.08	& 5.5	\\
	$S$, $\bullet$	& 0.12	& 5.5	\\
	\hline
	\multicolumn{3}{c}{Fig.~\ref{Results_psf_opo_AandS_N_v}, bottom}\\
	\hline
	$A$, $\times$	& -0.40	& 157	\\
	$A$, $\circ$	& -0.48	& 165	\\
	$A$, $\bullet$	& -0.55	& 173	\\
	\hline
	$S$, $\times$	& 0.84	& 38		\\
	$S$, $\circ$	& 0.69	& 94		\\
	$S$, $\bullet$	& 0.28	& 134	\\
	\hline
	\hline
      \end{tabular}
  &
      \begin{tabular}[t]{l r r}
        \hline
	\hline
	Reference		& $a$	& $b$	\\
	\hline
	\multicolumn{3}{c}{Fig.~\ref{AS_F6_90}}\\
	\hline
	$A$, $\Box$		& -0.13	& 38		\\
	$A$, $\circ$		& -0.5	& 152	\\
	$A$, $\vartriangle$	& -1.8	& 486	\\
	\hline
	$S$, $\Box$		& 0.006	& 1.03	\\
	$S$, $\circ$		& 0.45	& 65		\\
	$S$, $\vartriangle$	& 5.6		& 381   \\
        \hline
	\multicolumn{3}{c}{Fig.~\ref{AS_K2_90}}\\
	\hline
	$A$, $\Box$		& -0.11	& 47		\\
	$A$, $\circ$		& -0.16	& 73		\\
	$A$, $\vartriangle$	& -0.26	& 129	\\
	$A$, $\Diamond$	& -0.40	& 189	\\
	\hline
	$S$, $\Box$		& 0.06	& 1.5		\\
	$S$, $\circ$		& 0.25	& 4.0		\\
	$S$, $\vartriangle$	& 1.04	& 20		\\
	$S$, $\Diamond$	& 2.0		& 55		\\
	\hline
	\multicolumn{3}{c}{Fig.~\ref{AS_F6-K2-G2_90}}\\
	\hline
	$A$, $\Box$		& -0.13	& 38		\\
	$A$, $\circ$		& -0.13	& 41		\\
	$A$, $\vartriangle$	& -0.11	& 47		\\	
	$A$, $\blacksquare$		& -0.50	& 152	\\
	$A$, $\bullet$			& -0.48	& 165	\\
	$A$, $\blacktriangle$	& -0.40	& 189	\\
	\hline
	$S$, $\Box$		& 0.006	& 1.03	\\
	$S$, $\circ$		& 0.02	& 1.3		\\
	$S$, $\vartriangle$	& 0.06	& 1.5		\\
	$S$, $\blacksquare$		& 0.45	& 65		\\
	$S$, $\bullet$			& 0.69	& 94		\\
	$S$, $\blacktriangle$	& 2.0		& 55		\\
        \hline
	\hline
      	\end{tabular}
      \\
\end{tabular}
\end{center}
\end{table}

\end{appendix}

\end{document}